\documentclass[useAMS,usenatbib,a4paper]{mn2e}

\usepackage{amssymb}
\usepackage{amsmath}
\usepackage{graphicx}
\usepackage{subfigure}
\usepackage{booktabs}
\usepackage{txfonts}
\usepackage{ifpdf}
\ifpdf
  \usepackage[pdftex]{hyperref}
\else
  \usepackage[ps2pdf,colorlinks=true]{hyperref}
\fi

\newif\ifbw
\bwfalse

\newcommand{\eqn}[1]{(#1)}

\newcommand{\tbl}[1]{Table~#1}

\newcommand{\fig}[1]{Fig.~#1}

\newcommand{\sectn}[1]{Sec.~#1}

\newcommand{\eg}{\mbox{\it e.g.}}
\newcommand{\ie}{\mbox{\it i.e.}}

\newcommand{\cmb}{{CMB}}
\newcommand{\cmbtext}{{cosmic microwave background}}

\newcommand{\wmap}{{WMAP}}
\newcommand{\wmaptext}{{Wilkinson Microwave Anisotropy Probe}}

\newcommand{\healpix}{{\tt HEALPix}}

\newcommand{\lambdaarch}{{LAMBDA}}
\newcommand{\lambdaarchtext}{{Legacy Archive for Microwave Background Data Analysis}}

\newcommand{\fwhm}{{FWHM}}

\newcommand{\spcend}{\ensuremath{\:}}
\newcommand{\img}{\ensuremath{{\rm i}}}
\newcommand{\cconj}{\ensuremath{\ast}} 
\newcommand{\reals}{\ensuremath{\mathbb{R}}}

\newcommand{\ltwo}{\ensuremath{\mathrm{L}^2}}
\newcommand{\sphere}{\ensuremath{{\mathrm{S}^2}}}
\newcommand{\sothree}{\ensuremath{{\mathrm{SO}(3)}}}
\newcommand{\vect}[1]{\ensuremath{\mbox{\boldmath ${#1}$}}}

\newcommand{\dx}{\ensuremath{\mathrm{\,d}}}
\newcommand{\dmu}[1]{\ensuremath{\dx \Omega(#1)}}
\newcommand{\dmun}{\ensuremath{\dx \Omega}}

\newcommand{\sa}{\ensuremath{\vect{\hat{s}}}}
\newcommand{\saa}{\ensuremath{\theta}}
\newcommand{\sab}{\ensuremath{\varphi}}
\newcommand{\sas}{\ensuremath{\saa, \sab}}
\newcommand{\eul}{\ensuremath{\mathbf{\rho}}}
\newcommand{\euls}{\ensuremath{\eula, \eulb, \eulc}}
\newcommand{\eula}{\ensuremath{\alpha}}
\newcommand{\eulb}{\ensuremath{\beta}}
\newcommand{\eulc}{\ensuremath{\gamma}}
\newcommand{\el}{\ensuremath{\ell}}
\newcommand{\m}{\ensuremath{m}}
\newcommand{\n}{\ensuremath{n}}
\newcommand{\elm}{\ensuremath{{\el\m}}}

\newcommand{\elmax}{\ensuremath{{\el_{\rm max}}}}
\newcommand{\p}{\ensuremath{^\prime}}

\newcommand{\kron}[2]{\ensuremath{\delta_{{#1}{#2}}}}

\renewcommand{\exp}[1]{\ensuremath{{\rm e}^{#1}}}
\newcommand{\shfarg}[3]{\ensuremath{Y_{#1#2}({#3})}}
\newcommand{\shfargc}[3]{\ensuremath{Y_{#1#2}^\cconj({#3})}}

\newcommand{\shf}[2]{\ensuremath{Y_{#1#2}}}

\newcommand{\shc}[3]{\ensuremath{{#1}_{{#2}{#3}}}}

\newcommand{\leg}[2]{\ensuremath{P_{{#1}}({#2})}}
\newcommand{\aleg}[3]{\ensuremath{P_{#1}^{#2}({#3})}}

\newcommand{\sbessel}[1]{\ensuremath{j_{#1}}}

\newcommand{\dmatbig}{\ensuremath{D}}

\newcommand{\Dlmnp}{\ensuremath{ \dmatbig_{\m\n}^{\el}(\eul) }}

\newcommand{\dmatsmall}{\ensuremath{d}}

\newcommand{\rot}{\ensuremath{\mathcal{R}}}
\newcommand{\rotmat}{\ensuremath{\mathbf{R}}}

\newcommand{\wav}{\ensuremath{\psi}}

\newcommand{\suml}{\ensuremath{\sum_{\el=0}^{\infty}}}
\newcommand{\summ}{\ensuremath{\sum_{\m=-\el}^\el}}

\newcommand{\nside}{\ensuremath{{N_{\rm{side}}}}}
\newcommand{\npix}{\ensuremath{{N_{\rm{pix}}}}}

\newcommand{\order}{\ensuremath{\mathcal{O}}}

\newcommand{\sumlmb}{\ensuremath{\sum_{\el \m}}}

\newcommand{\sao}{\ensuremath{\sa_0}}
\newcommand{\saog}{\ensuremath{\sao^{\glob}}}
\newcommand{\sag}{\ensuremath{\sa^{\glob}}}
\newcommand{\sal}{\ensuremath{\sa^{\loc}}}
\newcommand{\sae}{\ensuremath{\sa^{\erth}}}
\newcommand{\san}{\ensuremath{\sa^{\nght}}}
\newcommand{\saoe}{\ensuremath{\sao^{\erth}}}
\newcommand{\saon}{\ensuremath{\sao^{\nght}}}

\newcommand{\salcoordfree}{\ensuremath{\vect{\sigma}}}

\newcommand{\blinemax}{\ensuremath{\|\vect{u}\|_{\rm max}}}

\newcommand{\blinel}{\ensuremath{\vect{u}}}

\newcommand{\blinee}{\ensuremath{\vect{u}^\erth}}
\newcommand{\vis}{\ensuremath{\mathcal{V}}}
\newcommand{\beam}{\ensuremath{A}}
\newcommand{\inten}{\ensuremath{I}}
\newcommand{\loc}{\ensuremath{{\rm l}}}
\newcommand{\glob}{\ensuremath{{\rm n}}}
\newcommand{\erth}{\ensuremath{{\rm e}}}
\newcommand{\nght}{\ensuremath{{\rm n}}}
\newcommand{\beaml}{\ensuremath{\beam^\loc}}

\newcommand{\intenl}{\ensuremath{\inten^\loc}}
\newcommand{\inteng}{\ensuremath{\inten^\glob}}
\newcommand{\intenn}{\ensuremath{\inten^\nght}}
\newcommand{\beammodintenl}{\ensuremath{\mbox{$\bigl( \beaml \cdot \intenl
    \bigr)$}}}
\newcommand{\beamhorzmodintenl}{\ensuremath{\mbox{$\bigl( \beaml \cdot
    \horizonl \cdot \intenl
    \bigr)$}}}

\newcommand{\horizon}{\ensuremath{H}}
\newcommand{\horizone}{\ensuremath{\horizon^\erth}}
\newcommand{\horizonl}{\ensuremath{\horizon^\loc}}

\newcommand{\nind}{\ensuremath{\hat{\vect{n}}_i}}
\newcommand{\none}{\ensuremath{\hat{\vect{n}}_1}}
\newcommand{\ntwo}{\ensuremath{\hat{\vect{n}}_2}}
\newcommand{\nthree}{\ensuremath{\hat{\vect{n}}_3}}

\newcommand{\eind}{\ensuremath{\hat{\vect{e}}_i}}
\newcommand{\eone}{\ensuremath{\hat{\vect{e}}_1}}
\newcommand{\etwo}{\ensuremath{\hat{\vect{e}}_2}}
\newcommand{\ethree}{\ensuremath{\hat{\vect{e}}_3}}
\newcommand{\uone}{\ensuremath{\hat{\vect{u}}_1}}
\newcommand{\utwo}{\ensuremath{\hat{\vect{u}}_2}}
\newcommand{\uthree}{\ensuremath{\hat{\vect{u}}_3}}

\newcommand{\saos}{\ensuremath{\saa_0, \sab_0}}
\newcommand{\roto}{\ensuremath{\rot_0}}
\newcommand{\rotofull}{\ensuremath{\rot(\sab_0,\saa_0,0)}}
\newcommand{\rotmato}{\ensuremath{\rotmat_0}}
\newcommand{\rott}{\ensuremath{\rot_t}}
\newcommand{\rotmatt}{\ensuremath{\rotmat_t}}

\newcommand{\lxvect}{\ensuremath{\vect{p}}}
\newcommand{\lx}{\ensuremath{p}}
\newcommand{\mx}{\ensuremath{q}}
\newcommand{\nimage}{\ensuremath{N_{\rm image}}}

\newcommand{\scalefun}{\ensuremath{\phi}}
\newcommand{\scal}{\ensuremath{j}}
\newcommand{\scalmax}{\ensuremath{J}}
\newcommand{\scalmin}{\ensuremath{J_0}}
\newcommand{\locat}{\ensuremath{k}}
\newcommand{\approxspace}{\ensuremath{V}}
\newcommand{\wavspace}{\ensuremath{W}}
\newcommand{\area}{\ensuremath{A}}
\renewcommand{\npix}{\ensuremath{N}}
\newcommand{\pixel}{\ensuremath{P}}
\newcommand{\wavtype}{\ensuremath{m}}
\newcommand{\acoeff}{\ensuremath{\lambda}}
\newcommand{\dcoeff}{\ensuremath{\gamma}}
\newcommand{\acoeffe}{\ensuremath{\eta}}
\newcommand{\dcoeffe}{\ensuremath{\delta}}

\title[Full-sky interferometry]
   {Simulating full-sky interferometric observations}

\author[McEwen \& Scaife]
  {J.~D.~McEwen\thanks{E-mail: mcewen@mrao.cam.ac.uk} and 
   A.~M.~M.~Scaife\\ 
  Astrophysics Group, 
      Cavendish Laboratory,  J.~J.~Thomson Avenue,
      Cambridge CB3 0HE, UK\\
}

\date{Accepted 3 July 2008. Received 30 June 2008; in original form 14
March 2008}
\pagerange{\pageref{sec:intro}--\pageref{lastpage}} 
\pubyear{2008}

\def\LaTeX{L\kern-.36em\raise.3ex\hbox{a}\kern-.15em
    T\kern-.1667em\lower.7ex\hbox{E}\kern-.125emX}

\begin{document}
\maketitle

\begin{abstract}
  Aperture array interferometers, such as that proposed for the Square
  Kilometre Array (SKA), will see the entire sky, hence the standard
  approach to simulating visibilities will not be applicable since it
  relies on a tangent plane approximation that is valid only for small
  fields of view.  We derive interferometric formulations in real,
  spherical harmonic and wavelet space that include contributions over
  the entire sky and do not rely on any tangent plane approximations.
  A fast wavelet method is developed to simulate the visibilities
  observed by an interferometer in the full-sky setting.  Computing
  visibilities using the fast wavelet method adapts to the sparse
  representation of the primary beam and sky intensity in the wavelet
  basis.  Consequently, the fast wavelet method exhibits superior
  computational complexity to the real and spherical harmonic space
  methods and may be performed at substantially lower computational
  cost, while introducing only negligible error to simulated
  visibilities.  Low-resolution interferometric observations are
  simulated using all of the methods to compare their performance,
  demonstrating that the fast wavelet method is approximately three
  times faster that the other methods for these low-resolution
  simulations.  The computational burden of the real and spherical
  harmonic space methods renders these techniques computationally
  infeasible for higher resolution simulations.  High-resolution
  interferometric observations are simulated using the fast wavelet
  method only, demonstrating and validating the application of this
  method to realistic simulations.  The fast wavelet method is
  estimated to provide a greater than ten-fold reduction in execution
  time compared to the other methods for these high-resolution
  simulations.
\end{abstract}


\section{Introduction}
\label{sec:intro}

The next generation of interferometers, such as the Square Kilometre
Array (SKA), will have to overcome a number of challenging imaging
issues.  Key among these is the question of how to deal with very
large fields of view, both in terms of the forward problem of
simulating observed visibilities and also in terms of the reverse
problem of reconstructing wide field images.  The
reverse wide field imaging problem has been tackled traditionally by
faceting the sky into a number of regions which are sufficiently small
that the standard tangent plane approximation to Fourier imaging is
possible \citep{cornwell:1992,greisen:2002}.  More recently,
\citet{cornwell:2005} have introduced the $w$-projection algorithm,
providing an order of magnitude speed improvement over facet-based
approaches.  The forward wide field imaging problem is an issue which
arises when observing with interferometric aperture arrays, such as
those proposed for the low frequency instrument of the SKA, and which
has yet to be resolved.


The final configuration and system design of the SKA are still under
active development. The final design will be dependent on the result
of simulations. Such simulations are created specifically to assess
different array configurations, whilst taking into account a range of
sky models and possible error contributions. These simulations are not
only important in determining the response of the interferometer to
the real sky but also in establishing the dynamic range of an
observation when faced with bright sources in the sidelobes of the
primary beam. As a consequence, simulating the response of an aperture
array interferometer correctly is as vital to the design studies of
such instruments as it is important for devising an effective method
of reducing the data when they finally arrive.

Such simulations are relatively simple for interferometers with small
fields of view.  Small fields can be well approximated as planes
tangent to the celestial sphere.  This approximation allows the
visibilities observed by an interferometer to be related to the
tangent plane image through the Fourier transform.  It is then
straightforward to simulate visibilities and reconstruct images.  In
the case of aperture arrays, which may see the entire hemisphere, the
operation is not so trivial. A tangent plane approximation is
obviously inappropriate for such geometries and hence the Fourier
transform can no longer be used to simulate observed visibilities. 

In this article we relax the small field of view assumption and tackle
interferometry when considering contributions over the entire sky.  We
derive full-sky interferometry formalisms using a number of
different representations, including representations in real,
spherical harmonic and wavelet spaces, and discuss the relative merits
of each approach.
Other authors have considered the spherical harmonic representation of
the interferometry integral relating the intensity of the sky to the
visibilities observed by an interferometer, predominantly in the
context of observations of the \cmbtext\ (\cmb).  \citet{white:1999}
relate the visibilities of observations of the \cmb\ to cosmological
quantities of interest, such as the angular power spectrum, through
contact with the spherical harmonics.  However, the flat tangent plane
approximation is still made and hence these results remain restricted
to small fields of view.  \citet{bunn:2007} extend the work of
\citet{white:1999} to larger fields of view by using mosaicing, making
the flat-sky approximation for each individual pointing used to
construct the mosaic.  \citet{ng:2001} was the first to present the
spherical harmonic representation of the interferometry integral while
including full-sky contributions, also in the context of computing the
angular power spectrum of observations of the \cmb.  \citet{ng:2001}
goes on to discuss implications of this result on properties of the
angular power spectrum recovered from interferometric observations.
In a separate piece of work, \citet{ng:2005} discusses the recovered
full-sky power spectra of \cmb\ experiments with asymmetric window
functions.  Although all of these works do address spherical harmonic
representations of interferometry, and in some cases include
contributions over the entire sky, the prevailing context of these
works is implications for the angular power spectrum recovered from
observations of the \cmb.  Relatively little attention has been paid
to full-sky interferometry formulations in the context of forward or
inverse wide field imaging, which pose important problems for next
generation interferometers.  The purpose of this article is to address
these issues, focusing particularly on the forward wide field imaging
problem.

The remainder of this article is organised as follows.  In
\sectn{\ref{sec:background}} we present the mathematical foundations
underlying results derived later in this paper, including a discussion
of harmonic analysis, wavelets and rotations on the sphere.  In
\sectn{\ref{sec:fsi}} we derive representations of the interferometric
visibility integral in real, spherical harmonic and wavelet space,
while including full-sky contributions.  The implications of these
results for forward and inverse wide field imaging are discussed.  In
\sectn{\ref{sec:sim}} we present resolution simulations of
full-sky interferometric observations using all of the methods
discussed in \sectn{\ref{sec:fsi}}.
Concluding remarks are made in \sectn{\ref{sec:conclusions}}.


\section{Mathematical preliminaries}
\label{sec:background}

Before presenting the formulation of interferometry on the full sky,
it is necessary to outline some mathematical preliminaries. We review
harmonic analysis and wavelets on the two-sphere \sphere, before
discussing rotations, which are represented by elements
of the rotation group \sothree.  By making all assumptions and
definitions explicit we hope to avoid any confusion over the
conventions adopted.


\subsection{Spherical harmonics}
\label{sec:background_sphere}

We consider the space of square integrable functions
$\ltwo(\sphere,\dmun)$ on the unit two-sphere $\sphere$, where
\mbox{$\dmu{\sa} = \sin\saa \dx\saa \dx\sab$} is the usual rotation
invariant measure on the sphere and $(\sas)$ denote the spherical
coordinates of $\sa \in \sphere$, with colatitude $\saa\in[0,\pi]$ and
longitude $\sab\in[0,2\pi)$.  A square integrable function on the
sphere $F \in \ltwo(\sphere,\dmun)$ may be represented by the
spherical harmonic expansion
\begin{displaymath}
F(\sa) = \sum_{\el=0}^\infty \sum_{\m=-\el}^\el \shc{F}{\el}{\m} \shfarg{\el}{\m}{\sa}
\spcend ,
\end{displaymath}
where the spherical harmonic coefficients are given by the usual
projection onto the spherical harmonic basis functions through the
inner product:
\begin{displaymath}
\shc{F}{\el}{\m} 
= 
\int_{\sphere}
F(\sa) \:
\shfargc{\el}{\m}{\sa}  
\dmu{\sa}
\spcend .
\end{displaymath}
The $\cconj$ denotes complex conjugation.  We adopt the Condon-Shortley
phase convention where the normalised spherical harmonics are defined
by \citep{varshalovich:1989}
\begin{displaymath}
\shfarg{\el}{\m}{\sa} = (-1)^\m \sqrt{\frac{2\el+1}{4\pi} 
\frac{(\el-\m)!}{(\el+\m)!}} \: 
\aleg{\el}{\m}{\cos\saa} \:
{\rm e}^{\img \m \sab}
\spcend ,
\end{displaymath}
where $\aleg{\el}{\m}{x}$ are the associated Legendre functions.
Using this normalisation the orthogonality of the spherical harmonic
functions reads
\begin{displaymath}
\label{eqn:shortho}
\int_\sphere
\shfarg{\el}{\m}{\sa} \:
\shfargc{\el\p}{\m\p}{\sa} 
\dmu{\sa}
=
\kron{\el}{\el\p} \kron{\m}{\m\p}
\spcend ,
\end{displaymath}
where $\delta_{ij}$ is the Kronecker delta function.  

We complete this section by noting two identities of which we will
make subsequent use.
Namely, we state the addition theorem for spherical
harmonics
\begin{equation}
\label{eqn:sh_addition}
\sum_{\m=-\el}^\el
\shfarg{\el}{\m}{\sa} \:
\shfargc{\el}{\m}{\sa\p}
=
\frac{2\el+1}{4\pi}
\leg{\el}{\sa \cdot \sa\p}
\end{equation}
and the Jacobi-Anger expansion of a plane wave
\begin{equation}
\label{eqn:jacobi_anger}
\exp{\img \vect{x} \cdot \vect{y}} =
\suml
(2\el+1) \:
\img^\el \:
\sbessel{\el}(\|\vect{x}\| \, \|\vect{y}\|) \:
\leg{\el}{\vect{\hat{x}} \cdot \vect{\hat{y}}}
\spcend ,
\end{equation}
where $\vect{x},\vect{y} \in \reals^3$, $\leg{\el}{\cdot}$ is the
Legendre function and $\sbessel{\el}(\cdot)$ is the spherical Bessel
function.

\subsection{Wavelets on the sphere}
\label{sec:background_wavelets}

Wavelets have proved useful in a wide range of applications due to
their ability to simultaneously resolve signal content in scale and
position.  In order to perform a wavelet analysis on the sphere, it is
necessary to extend the ordinary Euclidean wavelet framework to a
spherical manifold.
A number of attempts have been made to extend wavelets to the sphere.
Discrete second generation wavelets on the sphere that are based on a
multiresolution analysis have been developed
\citep{schroder:1995,sweldens:1996}.  Following the generic lifting
scheme proposed in these works, Haar wavelets on the sphere for
particular pixelisation schemes have also been developed
\citep{tenorio:1999,barreiro:2000}.  These discrete constructions
allow for the exact reconstruction of a signal from its wavelet
coefficients but they may not necessarily lead to a stable basis (see
\citet{sweldens:1997} and references therein).  Other authors have
focused on continuous wavelet methodologies on the sphere
\citep{freeden:1997a,freeden:1997b,holschneider:1996,torresani:1995,dahlke:1996,antoine:1998,antoine:1999,antoine:2002,antoine:2004,demanet:2003,wiaux:2005,sanz:2006,mcewen:2006:cswt2}.
Although signals can be reconstructed exactly from their wavelet
coefficients in these continuous methodologies in theory, the absence
of an infinite range of dilations precludes exact reconstruction in
practice.  Approximate reconstruction formula may be developed by
building discrete wavelet frames that are based on the continuous
methodology (\eg\ \citealt{bogdanova:2004}).  More recently, filter
bank wavelet methodologies that are essentially based on a continuous
wavelet framework have been developed for the axisymmetric
\citep{starck:2006} and directional \citep{wiaux:2007:sdw} cases.
These methodologies allow the exact reconstruction of a signal from
its wavelet coefficients in theory and in practice.
In the full-sky interferometry fomalism developed herein, we require
a wavelet analysis that allows the perfect reconstruction of a
function on the sphere from its wavelet coefficients.  Furthermore, we
also require an orthogonal wavelet analysis for reasons that will
become clear in \sectn{\ref{sec:fsi}}.  The only wavelet methodology
on the sphere that satisfies these requirements is the spherical Haar
wavelet (SHW) framework, which has the additional advantage of
simplicity and is also the most computationally efficient methodology.
We adopt the \healpix\footnote{\url{http://healpix.jpl.nasa.gov/}}
pixelisation of the sphere \citep{gorski:2005} for the implementation
of the SHW framework due to its hierarchical nature and ubiquitous use
in the astrophysical community.

The description of wavelets on the sphere given here is based largely
on the generic lifting scheme proposed by \citet{schroder:1995} and
also on the specific definition of Haar wavelets on a \healpix\
pixelised sphere proposed by \citet{barreiro:2000}.  However, our
discussion and definitions contain a number of notable differences to
those given by \citet{barreiro:2000} since we construct an orthonormal
Haar basis on the sphere and describe this in a multiresolution
setting.

We begin by defining a nested hierarchy of spaces as required for a
multiresolution analysis (see \citet{daubechies:1992} for a more
detailed discussion of multiresolution analysis).  Firstly, consider
the approximation space $\approxspace_\scal$ on the sphere \sphere,
which is a subset of the space of square integrable functions on the
sphere, \ie\ $\approxspace_\scal\subset\ltwo(\sphere,\dmun)$.  One may think
of $\approxspace_\scal$ as the space of piecewise constant functions
on the sphere, where the index $\scal$ corresponds to the size of the
piecewise constant regions.  As the resolution index $\scal$
increases, the size of the piecewise constant regions shrink, until in
the limit we recover $\ltwo(\sphere,\dmun)$ as $\scal\rightarrow\infty$.  If
the piecewise constants regions of $\sphere$ are arranged
hierarchically as $\scal$ increases, then one can construct the nested
hierarchy of approximation spaces
\begin{equation}
\approxspace_1 \subset \approxspace_2 \subset \cdots 
\subset \approxspace_\scalmax
\subset \ltwo(\sphere,\dmun)
\spcend ,
\label{eqn:space_hierarchy}
\end{equation}
where coarser (finer) approximation spaces correspond to a lower
(higher) resolution level $\scal$.  For each space
$\approxspace_\scal$ we define a basis with basis elements given by
the \emph{scaling functions}
$\scalefun_{\scal,\locat}\in\approxspace_\scal$, where the $\locat$ index
corresponds to a translation on the sphere.  Now, let us define
$\wavspace_\scal$ be the orthogonal complement of $\approxspace_\scal$
in $\approxspace_{\scal+1}$.
$\wavspace_\scal$ essentially provides a space for the representation
of the components of a function in $\approxspace_{\scal+1}$ that
cannot be represented in $\approxspace_\scal$, \ie\
$\approxspace_{\scal+1} = \approxspace_\scal \oplus \wavspace_\scal$.
For each space $\wavspace_\scal$ we define a basis with basis elements
given by the \emph{wavelets} $\wav_{\scal,\locat}\in\wavspace_\scal$.
The wavelet space $\wavspace_\scal$ encodes the difference (or
details) between two successive approximation spaces
$\approxspace_{\scal}$ and $\approxspace_{\scal+1}$.  By expanding the
hierarchy of approximation spaces, the highest level (finest) space
$\scal=\scalmax$, can then be represented by the lowest level
(coarsest) space $\scal=1$ and the differences between the
approximation spaces that are encoded by the wavelet spaces:
\begin{equation}
\label{eqn:multires}
V_\scalmax = V_1 \oplus \bigoplus_{\scal=1}^{\scalmax-1} W_j
\spcend .
\end{equation}

Let us now relate the generic description of multiresolution spaces
given above to the \healpix\ pixelisation.  The \healpix\ scheme
provides a hierarchical pixelisation of the sphere and hence may be
used to define the nested hierarchy of approximation spaces
explicitly.  The piecewise constant regions of the function spaces
$\approxspace_\scal$ discussed above now correspond to the pixels of
the \healpix\ pixelisation at the resolution associated with
$\approxspace_\scal$.  To make the association explicit, let
$\approxspace_\scal$ correspond to a \healpix\ pixelised sphere with
resolution parameter $\nside=2^{\scal-1}$ (\healpix\ spheres are
represented by the resolution parameter \nside, which is related to
the number of pixels in the pixelisation by $\npix=12\nside^2$).  In
the \healpix\ scheme, each pixel at level $\scal$ is subdivided into
four pixels at level $\scal+1$, and the nested hierarchy given by
\eqn{\ref{eqn:space_hierarchy}} is satisfied.  The number of pixels
associated with each space $\approxspace_\scal$ is given by
$\npix_\scal=12\times4^{\scal-1}$, where the area of each pixel is
given by $\area_\scal=4\pi/\npix_\scal=\pi / (3 \times 4^{\scal-1})$
(note that all pixels in a \healpix\ sphere at resolution $\scal$ have
equal area).  It is also useful to note that the number and area of
pixels at one level relates to adjacent levels through
$\npix_{\scal+1}=4\npix_{\scal}$ and $\area_{\scal+1}=\area_\scal / 4$
respectively.

\begin{figure*}
\centering
\includegraphics[height=80mm]{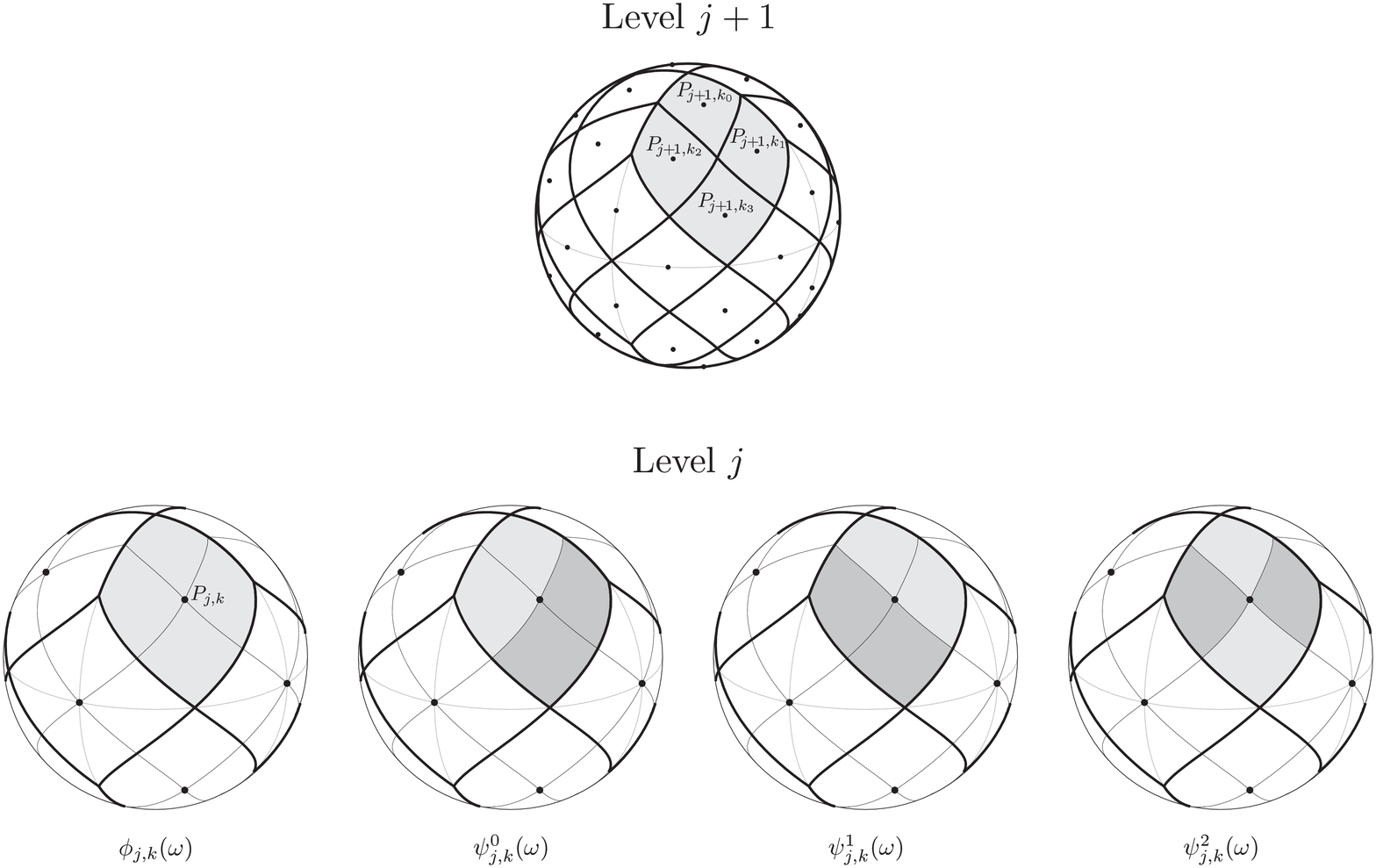}
\caption{Haar scaling function $ \scalefun_{\scal,\locat}(\sa)$ and
  wavelets $\wav_{\scal,\locat}^\wavtype(\sa)$.  Dark shaded regions
  correspond to negative constant values, light shaded regions
  correspond to positive constant values and unshaded regions
  correspond to zero.  The scaling function and wavelets at level
  \scal\ and position \locat\ are non-zero on pixel
  $\pixel_{\scal,\locat}$ only.  Pixel $\pixel_{\scal,\locat}$ at
  level $\scal$ is subdivided into four pixels at level $\scal+1$,
  which we label $\pixel_{\scal+1,\locat_0}$,
  $\pixel_{\scal+1,\locat_1}$, $\pixel_{\scal+1,\locat_2}$ and
  $\pixel_{\scal+1,\locat_3}$ as defined above.}
\label{fig:wavelets}
\end{figure*}

We are now in a position to define the scaling functions and wavelets
explicitly for the Haar basis on the nested hierarchy of \healpix\
spheres.  In this setting the index $\locat$ corresponds to the
position of pixels on the sphere, \ie\ for $\approxspace_\scal$ we get
the range of values $\locat=0,\cdots,\npix_\scal - 1$, and we let
$\pixel_{\scal,\locat}$ represent the region of the $\locat$th pixel of
a \healpix\ sphere at resolution $\scal$.  For the Haar basis, we
define the scaling function $\scalefun_{\scal,\locat}$ at level $\scal$
to be constant for pixel $\locat$ and zero elsewhere:
\begin{equation*}
\scalefun_{\scal,\locat}(\sa) \equiv
\begin{cases}
1/\sqrt{\area_\scal} & \sa \in \pixel_{\scal,\locat} \\
0                   & \text{elsewhere .}
\end{cases}
\end{equation*}
The non-zero value of the scaling function $1/\sqrt{\area_\scal}$ is
chosen to ensure that the scaling functions $\scalefun_{\scal,\locat}$
for $k=0,\cdots,\npix_\scal - 1$ do indeed define an orthonormal basis
for $\approxspace_\scal$.  Before defining the wavelets explicitly, we
fix some additional notation.  Pixel $\pixel_{\scal,\locat}$ at level $\scal$ is
subdivided into four pixels at level $\scal+1$, which we label
$\pixel_{\scal+1,\locat_0}$, $\pixel_{\scal+1,\locat_1}$,
$\pixel_{\scal+1,\locat_2}$ and $\pixel_{\scal+1,\locat_3}$, as
illustrated in \fig{\ref{fig:wavelets}}.  An orthonormal basis for the wavelet
space $\wavspace_\scal$, the orthogonal complement of
$\approxspace_\scal$, is then given by the following wavelets of type
$\wavtype=\{0,1,2\}$:
\begin{equation*}
\wav_{\scal,\locat}^0(\sa) \equiv \bigl [
\scalefun_{\scal+1,\locat_0}(\sa) -
\scalefun_{\scal+1,\locat_1}(\sa) +
\scalefun_{\scal+1,\locat_2}(\sa) -
\scalefun_{\scal+1,\locat_3}(\sa) \bigr ] / 2
\spcend ;
\end{equation*}
\begin{equation*}
\wav_{\scal,\locat}^1(\sa) \equiv \bigl [
\scalefun_{\scal+1,\locat_0}(\sa) +
\scalefun_{\scal+1,\locat_1}(\sa) -
\scalefun_{\scal+1,\locat_2}(\sa) -
\scalefun_{\scal+1,\locat_3}(\sa) \bigr ] / 2
\spcend ;
\end{equation*}
\begin{equation*}
\wav_{\scal,\locat}^2(\sa) \equiv \bigl [
\scalefun_{\scal+1,\locat_0}(\sa) -
\scalefun_{\scal+1,\locat_1}(\sa) -
\scalefun_{\scal+1,\locat_2}(\sa) +
\scalefun_{\scal+1,\locat_3}(\sa) \bigr ] / 2
\spcend .
\end{equation*}
We require three independent wavelet types to construct a complete
basis for $\wavspace_\scal$ since the dimension of
$\approxspace_{\scal+1}$ (given by $\npix_{\scal+1}$) is four times
larger than the dimension of $\approxspace_\scal$ (the approximation
function provides the fourth component).  The Haar scaling functions
and wavelets defined here on the sphere are illustrated in
\fig{\ref{fig:wavelets}}.

Let us check that the scaling functions and wavelets satisfy the
requirements for an orthonormal multiresolution analysis as outlined
previously.  We require $\wavspace_\scal$ to be orthogonal to
$\approxspace_\scal$, \ie\ we require
\begin{equation*}
\int_\sphere 
\scalefun_{\scal,\locat}(\sa) \:
\wav_{\scal,\locat^\prime}^\wavtype(\sa) 
\dmu{\sa} = 0
\spcend.
\end{equation*}
This is always satisfied since for $\locat^\prime\neq\locat$ the
scaling function and wavelet do not overlap and so the integrand is
zero always, and for $\locat^\prime=\locat$ we find
\begin{equation*}
\int_\sphere 
\scalefun_{\scal,\locat}(\sa) \: \wav_{\scal,\locat}^\wavtype(\sa)
\dmu{\sa}
\propto
\int_\sphere 
\wav_{\scal,\locat}^\wavtype(\sa) 
\dmu{\sa}
= 0
\spcend .
\end{equation*}
We also require $\wavspace_\scal$ to be orthogonal to
$\wavspace_{\scal^\prime}$ for all $\scal$ and $\scal^\prime$.  Again,
if the basis functions do not overlap (\ie\ $\locat\neq\locat^\prime$)
then this requirement is satisfied automatically, and if they do (\ie\
$\locat=\locat^\prime$) then the wavelet at the finer level
$\scal^\prime>\scal$ will always lie within a region of the wavelet at
level $\scal$ with constant value, and consequently
\begin{equation*}
\int_\sphere 
\wav_{\scal,\locat}^\wavtype(\sa) \: \wav_{{\scal^\prime},{\locat^\prime}}^{\wavtype^\prime}(\sa)
\dmu{\sa}
\propto
\int_\sphere 
\wav_{\scal^\prime,\locat^\prime}^{\wavtype^\prime}(\sa)
\dmu{\sa}
= 0
\spcend .
\end{equation*}
Finally, to ensure that we have constructed an orthonormal wavelet
basis for $\wavspace_\scal$, we check the orthogonality of all
wavelets at level $\scal$:
\begin{equation*}
\int_\sphere 
\wav_{\scal,\locat}^\wavtype(\sa) \:
\wav_{{\scal},{\locat^\prime}}^{\wavtype^\prime}(\sa)
\dmu{\sa}
= \kron{\wavtype}{\wavtype^\prime}
\kron{\locat}{\locat^\prime} \:
\Biggl(\frac{1}{2\sqrt{\area_{\scal+1}}} \Biggr)^2 \: \area_\scal
=\kron{\wavtype}{\wavtype^\prime}
\kron{\locat}{\locat^\prime}
\spcend ,
\end{equation*}
where for $\wavtype\neq\wavtype^\prime$ the positive and negative
regions of the integrand cancel exactly and for
$\locat\neq\locat^\prime$ the wavelets do not overlap and so the
integrand is zero always.  Note that in the previous expression the
final $\area_\scal$ term arises from the area element.  The
Haar approximation and wavelet spaces that we have constructed
therefore satisfy the requirements of an orthonormal multiresolution
analysis on the sphere.  Although the orthogonal nature of these
spaces is important, a different normalisation could be chosen.  It is
now possible to define the analysis and synthesis of a function on the
sphere in this Haar wavelet multiresolution framework.

\begin{figure*}
\centering
\includegraphics[height=120mm]{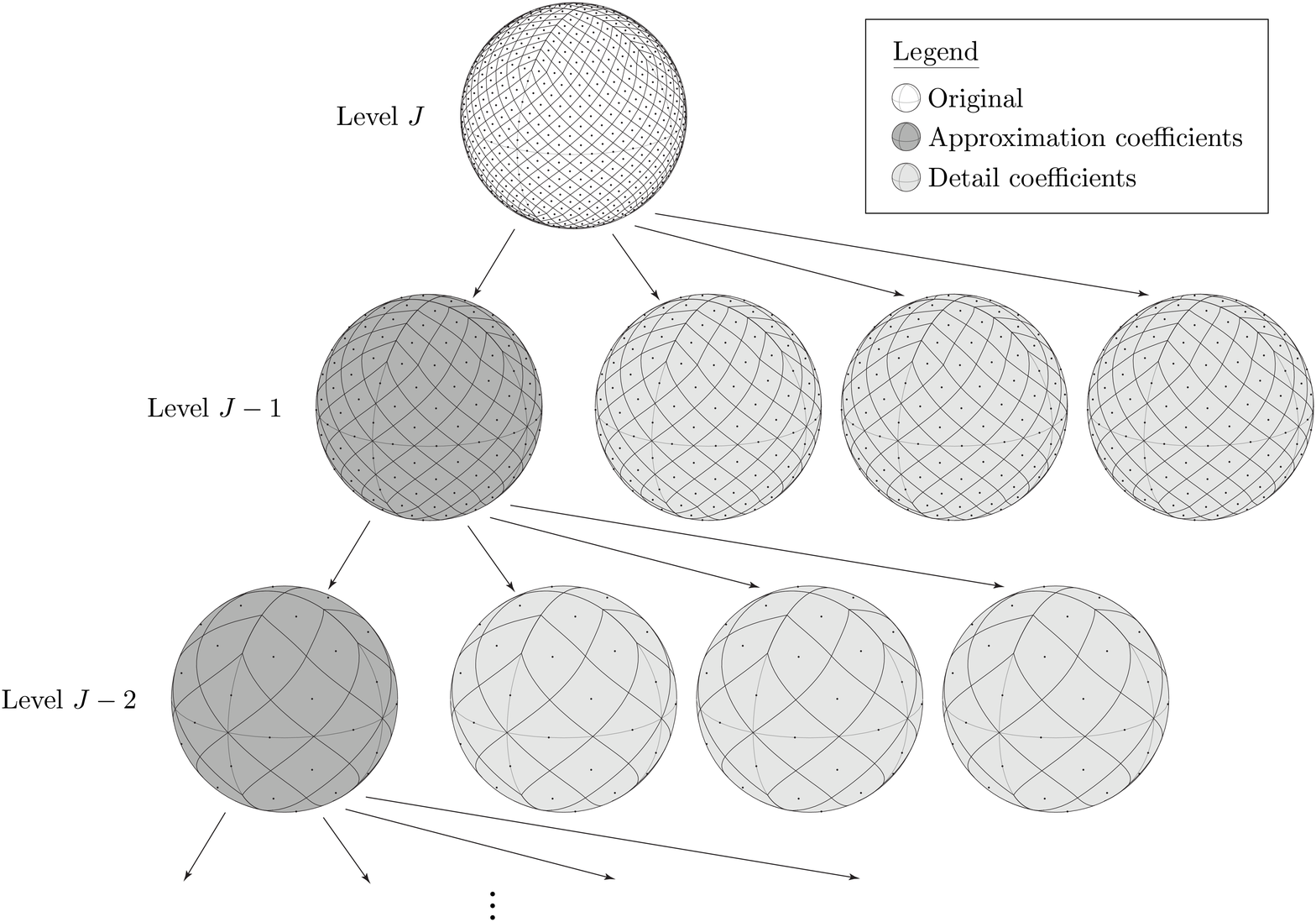}
\caption{Haar multiresolution decomposition.  Starting at the finest
  level $\scalmax$ (the original sphere), the approximation and
  detail coefficients at level $\scalmax-1$ are computed.  This
  procedure is repeated to decompose the approximation coefficients at
  level $\scalmax-1$ (\ie\ the approximation function
  $F_{\scalmax-1}$), into approximation and detail coefficients at the
  coarser level $\scalmax-2$.  Repeating this procedure continually,
  one recovers the multiresolution representation of $F_\scalmax$ in
  terms of the coarsest level approximation $F_{\scalmin}$ and all of
  the detail coefficients.}
\label{fig:multiresolution}
\end{figure*}

The decomposition of a function defined on a \healpix\ sphere at
resolution $\scalmax$, \ie\ $F_\scalmax\in\approxspace_\scalmax$, into
its wavelet and scaling coefficients proceeds as follows.  Consider an
intermediate level $\scal+1<\scalmax$ and let $F_{\scal+1}$ be the
approximation of $F_\scalmax$ in $\approxspace_{\scal+1}$.  The
scaling coefficients at the coarser level $\scal$ are given by the
projection of $F_{\scal+1}$ onto the scaling functions
$\scalefun_{\scal,\locat}$:
\begin{align}
\acoeff_{\scal,\locat} &\equiv \int_\sphere
F_{\scal+1}(\sa)  \:
\scalefun_{\scal,\locat}(\sa) \dmu{\sa} \nonumber \\
& = \bigl (
\acoeff_{\scal+1,\locat_0} +
\acoeff_{\scal+1,\locat_1} +
\acoeff_{\scal+1,\locat_2} +
\acoeff_{\scal+1,\locat_3}
\bigr ) \sqrt{\area_\scal}/4
\spcend , \nonumber
\end{align}
where we call $\acoeff_{\scal,\locat}$ the \emph{approximation coefficients}
since they define the approximation function
$F_\scal\in\approxspace_\scal$.  At the finest level $\scalmax$, we
naturally associate the function values of $F_\scalmax$ with the
approximation coefficients of this level.  The wavelet coefficients at
level $\scal$ are given by the projection of $F_{\scal+1}$ onto the wavelets
$\wav_{\scal,\locat}^\wavtype$:
\begin{equation*}
\dcoeff_{\scal,\locat}^\wavtype \equiv
\int_\sphere
F_{\scal+1}(\sa) \:
\wav_{\scal,\locat}^\wavtype(\sa) \dmu{\sa}
\spcend ,
\end{equation*}
giving
\begin{equation*}
\dcoeff_{\scal,\locat}^0 = \bigl (
\acoeff_{\scal+1,\locat_0} -
\acoeff_{\scal+1,\locat_1} +
\acoeff_{\scal+1,\locat_2} -
\acoeff_{\scal+1,\locat_3}
\bigr ) \sqrt{\area_\scal}/4
\spcend ,
\end{equation*}
\begin{equation*}
\dcoeff_{\scal,\locat}^1 = \bigl (
\acoeff_{\scal+1,\locat_0} +
\acoeff_{\scal+1,\locat_1} -
\acoeff_{\scal+1,\locat_2} -
\acoeff_{\scal+1,\locat_3}
\bigr ) \sqrt{\area_\scal}/4
\spcend 
\end{equation*}
and
\begin{equation*}
\dcoeff_{\scal,\locat}^2 = \bigl (
\acoeff_{\scal+1,\locat_0} -
\acoeff_{\scal+1,\locat_1} -
\acoeff_{\scal+1,\locat_2} +
\acoeff_{\scal+1,\locat_3}
\bigr ) \sqrt{\area_\scal}/4
\spcend ,
\end{equation*}
where we call $\dcoeff_{\scal,\locat}^\wavtype$ the \emph{detail (or
wavelet) coefficients} of type $\wavtype$.
Starting from the finest level $\scalmax$, we compute the
approximation and detail coefficients at level $\scalmax-1$ as
outlined above.  We then repeat this procedure to decompose the
approximation coefficients at level $\scalmax-1$ (\ie\ the
approximation function $F_{\scalmax-1}$), into approximation and
detail coefficients at the coarser level $\scalmax-2$.  Repeating this
procedure continually, we recover the multiresolution representation
of $F_\scalmax$ in terms of the coarsest level approximation $F_1$ and
all of the detail coefficients, as specified by
\eqn{\ref{eqn:multires}} and illustrated in \fig{\ref{fig:multiresolution}}.  In
general it is not necessary to continue the multiresolution
decomposition down to the coarsest level $\scal=1$; one may choose to
stop at the intermediate level $\scalmin$, where $1\leq \scalmin <
\scalmax$.

The function $F_\scalmax\in\approxspace_\scalmax$ may then be
synthesised from its approximation and detail coefficients.  Due to
the orthogonal nature of the Haar basis, the approximation
coefficients at level $\scal+1$ may be reconstructed from the weighted
expansion of the scaling function and wavelets at the coarser level $\scal$,
where the weights are given by the approximation and detail
coefficients respectively.  Writing this expansion explicitly, the
approximation coefficients at level $\scal+1$ are given in terms of the
approximation and detail coefficients of the coarser level $\scal$:
\begin{equation*}
\acoeff_{\scal+1,\locat_0} = \bigl (
\acoeff_{\scal,\locat} +
\dcoeff_{\scal,\locat}^0 +
\dcoeff_{\scal,\locat}^1 +
\dcoeff_{\scal,\locat}^2
\bigr ) / \sqrt{\area_\scal}
\spcend ;
\end{equation*}
\begin{equation*}
\acoeff_{\scal+1,\locat_1} = \bigl (
\acoeff_{\scal,\locat} -
\dcoeff_{\scal,\locat}^0 +
\dcoeff_{\scal,\locat}^1 -
\dcoeff_{\scal,\locat}^2
\bigr ) / \sqrt{\area_\scal}
\spcend ;
\end{equation*}
\begin{equation*}
\acoeff_{\scal+1,\locat_2} = \bigl (
\acoeff_{\scal,\locat} +
\dcoeff_{\scal,\locat}^0 -
\dcoeff_{\scal,\locat}^1 -
\dcoeff_{\scal,\locat}^2
\bigr ) / \sqrt{\area_\scal}
\spcend ;
\end{equation*}
\begin{equation*}
\acoeff_{\scal+1,\locat_3} = \bigl (
\acoeff_{\scal,\locat} -
\dcoeff_{\scal,\locat}^0 -
\dcoeff_{\scal,\locat}^1 +
\dcoeff_{\scal,\locat}^2
\bigr ) / \sqrt{\area_\scal}
\spcend .
\end{equation*}
Repeating this procedure from level $\scal=\scalmin$ up to
$\scal=\scalmax$, one finds that the signal
$F_\scalmax\in\approxspace_\scalmax$ may be written as the scaling
function and wavelet expansion
\begin{equation}
F_\scalmax(\sa) =
\sum_{\locat=0}^{\npix_{\scalmin} - 1} 
\acoeff_{\scalmin,\locat} \: \scalefun_{\scalmin,\locat}(\sa)
+
\sum_{\scal=\scalmin}^{\scalmax-1}
\sum_{\locat=0}^{\npix_\scal - 1}
\sum_{\wavtype=0}^{2}
\dcoeff_{\scal,\locat}^\wavtype \:
\wav_{\scal,\locat}^\wavtype(\sa)
\spcend .
\end{equation}

\subsection{Rotations}
\label{sec:background_rot}

Rotations $\rot$ on the sphere are characterised by the elements of
the rotation group $\sothree$, which we parameterise in terms of the
three Euler angles \mbox{$\rho=(\euls)\in \sothree$}, where
$\eula\in[0,2\pi)$, $\eulb\in[0,\pi]$ and $\eulc\in[0,2\pi)$. The
rotation of a coordinate vector $\sa$ by $\rot(\eul)$ may be
represented by multiplication of the Cartesian coordinate with the
3$\times$3 rotation matrix $\rotmat(\eul)$.  We adopt the $zyz$ Euler
convention corresponding to the rotation of a physical body in a
fixed coordinate system about the $z$, $y$ and $z$ axes by
$\eulc$, $\eulb$ and $\eula$ respectively, \ie\ $\rotmat(\eul) =
\rotmat_z(\eula) \rotmat_y(\eulb) \rotmat_z(\eulc)$, where
$\rotmat_z(\vartheta)$ and $\rotmat_y(\vartheta)$ are rotation
matrices representing rotations by $\vartheta$ about the $z$ and $y$
axis respectively.  The inverse rotation is given by
$\rotmat^{-1}(\eul) = \rotmat_z(-\eulc) \rotmat_y(-\eulb)
\rotmat_z(-\eula)$.  We define the rotation of a function $F \in
\ltwo(\sphere,\dmun)$ on the sphere by
\begin{equation}
\label{eqn:rot}
\bigl(\rot(\eul) F\bigr)(\sa) = F\bigl(\rotmat^{-1}(\eul) \sa\bigr)
\spcend .
\end{equation}

It is also useful to characterise the rotation of a function on the
sphere in harmonic space.  The Wigner \mbox{$\dmatbig$-functions}
$\Dlmnp$ provide the irreducible unitary representation of the
rotation group $\sothree$.  
For our purpose, we merely consider
the Wigner \mbox{$\dmatbig$-functions} to represent the
rotation of a function on the sphere in harmonic space.
The rotation of a spherical harmonic basis function may be represented
by a sum of weighted harmonics of the same \el\ \citep{brink:1999}:
$
\bigl(\rot(\eul)\shf{\el}{\m}\bigr)(\sa) = 
\sum_{\n=-\el}^{\el} 
\dmatbig_{\n\m}^{\el}(\eul) \: \shfarg{\el}{\n}{\sa}
$.
It is then trivial to show that the harmonic
coefficients of a rotated function are related to the coefficients of
the original function by
\begin{equation}
\label{eqn:shrot:2}
\bigl(\rot(\eul) F \bigr)_\elm = 
\sum_{\n=-\el}^{\el} 
\dmatbig_{\m\n}^{\el}(\eul) \: 
F_{\el\n}
\spcend .
\end{equation}
%
For computational purposes, the Wigner functions may be decomposed as
$
\dmatbig_{\m\n}^{\el}(\euls)
= {\rm e}^{-\img \m\eula} \:
\dmatsmall_{\m\n}^\el(\eulb) \:
{\rm e}^{-\img \n\eulc}
$,  
where the real polar \dmatsmall-functions are defined by \citet{varshalovich:1989}.
Recursion formulae exist to compute the Wigner \dmatsmall-functions
rapidly \citep{risbo:1996}.

\section{Full-sky interferometry}
\label{sec:fsi}

We formulate full-sky interferometry in this section in real,
spherical harmonic and SHW spaces.  Since we treat interferometry in
the full-sky setting it is necessary to be explicit about the
coordinate systems used and the relations between them.  After
discussing the various coordinate systems that we use, we describe
visibility computation and image reconstruction for the representation
of interferometry in each space, highlighting the relative merits of
each representation.  These ideas are then extended to incorporate
horizon occlusion and primary beam functions that depend on the
interferometer pointing direction.


\subsection{Coordinate systems}
\label{sec:fsi_coord}

\begin{figure}
\centering
\includegraphics[width=50mm]{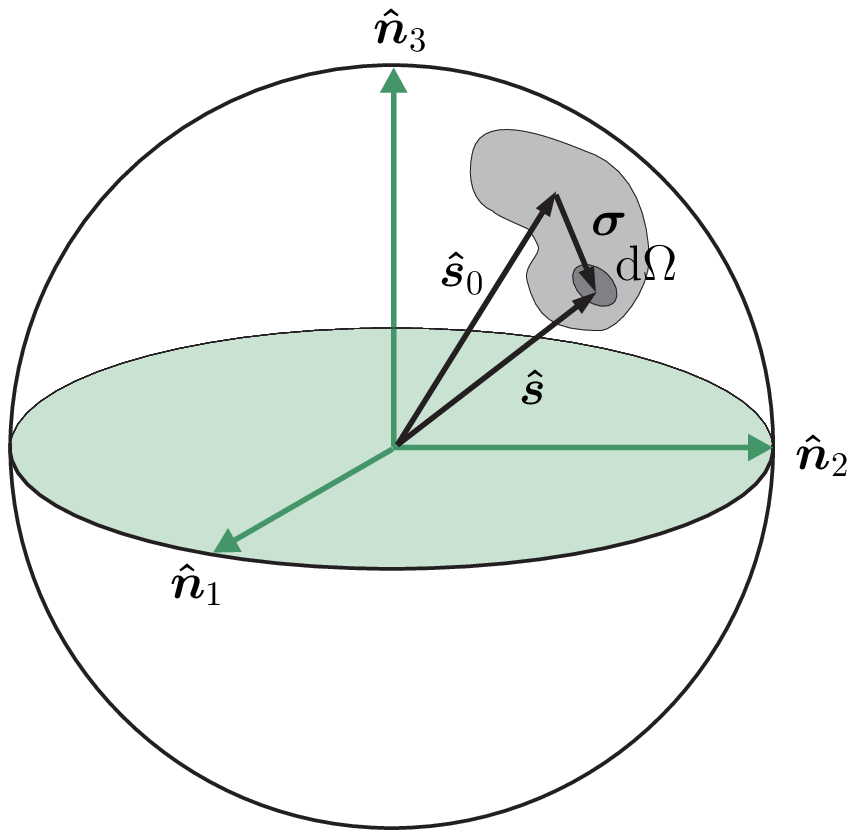}
\caption{Geometry of an observation of an extended source centred at
  the interferometer pointing direction $\sao$.  The area element
  $\dmun$ represents the contribution to the visibility integral from
  point $\sa$.  The full visibility is obtained by summing all such
  contributions over the sky.  Note that the sky has been mapped onto
  the unit celestial sphere, hence $\sa$ and $\sao$ are unit vectors.}
\label{fig:infer_obs}
\end{figure}

The complex visibility measured by an interferometer is given by the
coordinate free definition \citep{thompson:2001}
\begin{equation}
\vis(\blinel) = \int_\sphere 
\beam(\salcoordfree) 
\inten(\salcoordfree)
\exp{- \img 2 \pi \blinel \cdot \salcoordfree}
\dx \Omega
\spcend ,
\end{equation}
where $\salcoordfree=\sa-\sao$; the area element $\dx \Omega$ and the
vectors $\salcoordfree$, $\sa$ and $\sao$ are defined in
\fig{\ref{fig:infer_obs}}.  We assume that the sky is mapped onto the
unit celestial sphere, hence all vectors on the sphere are unit
vectors.  The vector $\blinel$ is related to the baseline of the
interferometer (and is defined explicitly later), the function $\beam$
defines the primary beam of the interferometer and the function
$\inten$ defines the intensity of the sky at position $\salcoordfree$.
In this coordinate free definition of visibility, $\salcoordfree$ is
essentially the representation of $\sa$ in a coordinate system centred
on $\sao$.  The translation $\salcoordfree=\sa-\sao$ represents the
transformation between the global coordinate frame of $\sa$ and the
local coordinate frame of $\salcoordfree$.  In general, one can
transform vectors between global coordinates and local coordinates
relative to $\sao$, through a rotation by $\sao$.  We now formalise
this notion.

\begin{figure}
\centering
\includegraphics[width=50mm]{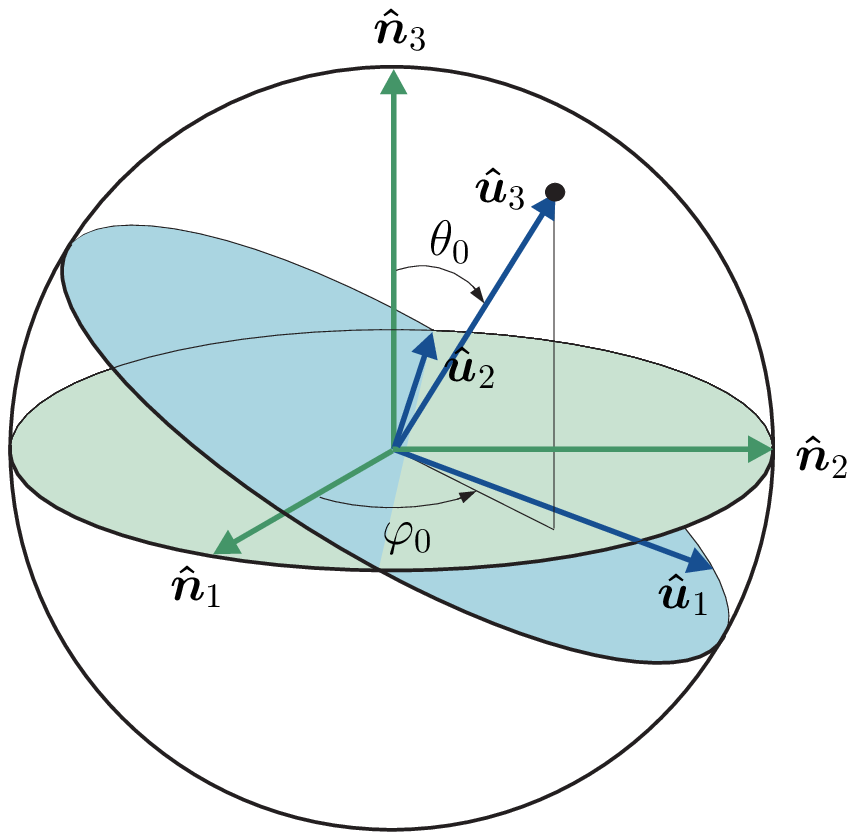}
\caption{Rotation $\roto$ mapping global coordinates of the celestial sky
  defined in the frame $\{\none,\ntwo,\nthree\}$ to local coordinates
  defined in the frame $\{\uone,\utwo,\uthree\}$, where $\uthree$ is
  aligned with $\sao$.  Local coordinates are related to global
  coordinates by $\sal=\rotmato^{-1} \sag$.}
\label{fig:infer_coord}
\end{figure}

Let us define two coordinate frames.  The global coordinate frame of
the celestial sky is defined by the right handed set of unit vectors
$\{\none,\ntwo,\nthree\}$.  The local coordinate frame relative to
$\sao$ is defined by the right handed set of unit vectors
$\{\uone,\utwo,\uthree\}$, where $\uthree$ aligns with $\sao$.  These
two coordinate frames are related by the rotation
$\roto\equiv\rotofull$, where $(\saos)$ are the spherical coordinates
of $\sao$ in the global coordinate frame.  The corresponding
$3\times3$ rotation matrix $\rotmato\equiv\rotmat(\sab_0,\saa_0,0)$
transforms a vector between local and global coordinates.  For
example, $\sao$ in global coordinates corresponds to $\uthree$ in
local coordinates (by definition) and the two vectors are related by
$\sao=\rotmato \uthree$.  In general, local coordinates are related to
global coordinates by $\sal=\rotmato^{-1} \sag$, where the
superscripts $\loc$ and $\glob$ denote local and global coordinates
respectively; henceforth all vectors and functions contain an $\loc$
or $\glob$ superscript to denote their coordinate frame.  The third
Euler angle of the rotation mapping local to global coordinates
specifies an azimuthal rotation around $\sao$, which is in general
arbitrary but fixed.  Without loss of generality we set this third
Euler angle to zero.  \fig{\ref{fig:infer_coord}} illustrates the
rotation that transforms between local and global coordinates.

Returning to the visibility function, we may now represent this in the
coordinate frames defined above.  The beam function is most naturally
represented in local coordinates relative to the pointing direction
$\saog$.  We denote this function by
$\beaml(\sal)\in\ltwo(\sphere,\dmun)$.  The source intensity function
is most naturally represented in global coordinates and is denoted by
$\inteng(\sag)\in\ltwo(\sphere,\dmun)$.  We may convert a function
$F^\glob$ in global coordinates to a corresponding function $F^\loc$
in local coordinates through the rotation $\roto$:
\begin{displaymath}
F^\glob(\sag) = 
F^\glob(\rotmato\sal) = 
(\roto^{-1}F^\glob)(\sal) = 
F^\loc(\sal) 
\spcend ,
\end{displaymath}
\ie\ $F^\loc=\roto^{-1} F^\glob$.
In practice, sampled functions on the sphere may be rotated in real
space through \eqn{\ref{eqn:rot}} or alternatively, and more accurately
(since pixelisation artifacts are eliminated), in harmonic space
through \eqn{\ref{eqn:shrot:2}}.
In this coordinate setting the visibility integral may be written
\begin{equation*}
\vis(\blinel) = \int_\sphere 
\beaml(\sal) \:
\inteng(\sag) \:
\exp{- \img 2 \pi \blinel \cdot \sal}
\dmu{\sal}
\spcend .
\end{equation*}
When dealing with the visibility integral it is more convenient to
represent all functions in the same coordinate frame.  In local
coordinates the visibility integral becomes
\begin{align}
\label{eqn:fsi_local}
\vis(\blinel) &= 
\int_\sphere 
\beaml(\sal) \:
(\roto^{-1} \inteng)(\sal) \:
\exp{- \img 2 \pi \blinel \cdot \sal}
\dmu{\sal} \nonumber \\
&=
\int_\sphere 
\beaml(\sal) \:
\intenl(\sal) \:
\exp{- \img 2 \pi \blinel \cdot \sal}
\dmu{\sal}
\spcend .
\end{align} 
Here $\blinel$ is the interferometer baseline in local coordinates (an
explicit expression for $\blinel$ is deferred until the more general
formulation presented in \sectn{\ref{sec:fsi_horizon}}).
The expression given by \eqn{\ref{eqn:fsi_local}} is the familiar
interferometric visibility integral, however in discussing this
integral in the full-sky setting it has been necessary to make the use
of different coordinate systems explicit.  To compute visibilities
when including contributions over the entire sky, one could simply
evaluate \eqn{\ref{eqn:fsi_local}} by using an appropriate quadrature
rule on the sphere.  The complexity of evaluating this integral
directly for a single baseline $\blinel$ scales as $\order(\npix)$
(recall that $\npix$ is the number of pixels contained in the
pixelisation of the sphere).  Low-resolution simulations of full-sky
interferometric observations are computed in this manner in
\sectn{\ref{sec:sim}}.

We also define the coordinate system $\lxvect=(\lx,\mx)$ on the
tangent plane of the celestial sphere at the pointing direction.  This
coordinate system is used to define the image plane when
reconstructing images on small patches using the standard Fourier
transform approach and is obviously a local coordinate frame.  It is
related to the spherical coordinates $(\saa^\loc,\sab^\loc)$ of the
local $\{\uone,\utwo,\uthree\}$ frame by $\lx=\sin\saa^\loc
\cos\sab^\loc$ and $\mx=\sin\saa^\loc \sin\sab^\loc$.


\subsection{Spherical harmonic space representation}

It is more general and accurate to compute the visibility integral in
harmonic space since this avoids the need for any quadrature rule on
the sphere, which would necessarily be pixelisation dependent and not
always exact.  Moreover, rotations can be performed through
\eqn{\ref{eqn:shrot:2}} in harmonic space more accurately than they can
be performed in real space.  We derive here the spherical harmonic
representation of the visibility integral and discuss the
implications of this representation for computing visibilities and for
image reconstruction.

\subsubsection{Computing visibilities}
\label{sec:fsi_vis}

Consider the beam-modulated source intensity function
$\beammodintenl(\sal) = \beaml(\sal) \intenl(\sal)$.  Substituting the
spherical harmonic expansion of this function into the local
coordinate visibility integral given by \eqn{\ref{eqn:fsi_local}}, we
obtain
\begin{equation}
\label{eqn:fsi_harm1}
\vis(\blinel) =
\sumlmb
\shc{\beammodintenl}{\el}{\m} \int_\sphere 
\exp{- \img 2 \pi \blinel \cdot \sal} \:
\shfarg{\el}{m}{\sal}
\dmu{\sal}
\spcend ,
\end{equation}
where
$\shc{\beammodintenl}{\el}{\m}$ are the spherical harmonic
coefficients of the beam-modulated intensity function.
We assume that the beam-modulated intensity function is band-limited
at $\elmax$ so that all higher frequency harmonic coefficients are
zero, \ie\ $\shc{\beammodintenl}{\el}{\m}=0$, $\forall \el>\elmax$.
In this case sums over the harmonic index $\el$ may be truncated at
$\elmax$.  Here, and subsequently, we use the shorthand notation
$\sumlmb\equiv\sum_{\el=0}^{\elmax}\summ$.  The integral contained in
\eqn{\ref{eqn:fsi_harm1}} may be evaluated analytically.  Noting the
addition theorem for spherical harmonics and the Jacobi-Anger
expansion of a plane wave given by \eqn{\ref{eqn:sh_addition}} and
\eqn{\ref{eqn:jacobi_anger}} respectively, we find
\begin{equation}
\label{eqn:wave_sh}
\exp{ \img 2 \pi \blinel \cdot \sal} =
4\pi
\sumlmb
\img^\el \:
\sbessel{\el}(2\pi \|\blinel\|) \:
\shfargc{\el}{\m}{\hat{\blinel}} \:
\shfarg{\el}{\m}{\sal}
\spcend .
\end{equation}
Using this result, the integral contained in
\eqn{\ref{eqn:fsi_harm1}} becomes
\begin{equation*}
\int_\sphere 
\exp{- \img 2 \pi \blinel \cdot \sal} \:
\shfarg{\el}{m}{\sal}
\dmu{\sal} 
=4\pi \:
(-\img)^\el \:
\sbessel{\el}(2\pi \|\blinel\|) \:
\shfarg{\el}{\m}{\vect{\hat{\blinel}}}
\spcend ,
\end{equation*}
where we have noted the orthogonality of the spherical harmonics.
The harmonic representation of the visibility function then reads
\begin{equation}
\label{eqn:fsi_harm}
\vis(\blinel) =
4\pi
\sumlmb 
(-\img)^\el \:
\sbessel{\el}(2\pi \|\blinel\|) \:
\shfarg{\el}{\m}{\vect{\hat{\blinel}}} \:
\shc{\beammodintenl}{\el}{\m} 
\spcend .
\end{equation}
This expression has been derived independently by \citet{ng:2001} to
study properties of the angular power spectrum recovered from
interferometric observations of the \cmb, however slightly different
definition are adopted to those used here.

Computing visibilities using \eqn{\ref{eqn:fsi_harm}} ensures that
full-sky contributions from the beam-modulated intensity function to
the visibility integral are incorporated, thus allowing
for arbitrarily wide fields of view and beam sizes.
%
%
The complexity of evaluating \eqn{\ref{eqn:fsi_harm}} for a single
baseline $\blinel$ scales as $\order(\elmax^2)$, where
$\order(\elmax^2) \sim \order(\npix)$ (as discussed in more detail in
\sectn{\ref{sec:sim_practical}}).  Low-resolution simulations of full-sky
interferometric observations are computed in this manner in
\sectn{\ref{sec:sim}}.

\subsubsection{Image reconstruction}
\label{sec:fsi_recon}

In theory one may recover a full-sky synthesised image from the
spherical harmonic representation of the visibility function by
integration over a spherical surface in $\reals^3$.  We describe this
approach before discussing its limitations.

It is possible to recover the beam-modulated intensity function by
taking the spherical harmonic transform of the visibility function
over a spherical surface at radius $\|\blinel\|$.  By substituting the
harmonic representation of the visibility function given by
\eqn{\ref{eqn:fsi_harm}} into the spherical harmonic transform of the
visibility one finds
\begin{equation}
\int_\sphere
\vis(\blinel) \:
\shfargc{\el}{\m}{\hat{\blinel}} \:
\dmu{\hat{\blinel}} 
= 
4\pi \:
(-\img)^\el \:
\sbessel{\el}(2\pi \|\blinel\|) \:
\shc{\beammodintenl}{\el}{\m} 
 ,
\label{eqn:fsi_syn}
\end{equation}
where we have noted the orthogonality of the spherical harmonics.  In
theory, \eqn{\ref{eqn:fsi_syn}} can be used to recover the harmonic
coefficients of the beam-modulated intensity, from which the real
space function can be reconstructed easily.  In this framework, one
recovers the beam-modulated intensity on the entire sky, arbitrarily
far from the interferometer pointing direction.  However, there are a
number of limitations of this approach that render it infeasible in
practice.

To recover the harmonic coefficients $\shc{\beammodintenl}{\el}{\m}$
from \eqn{\ref{eqn:fsi_syn}} we require that $\sbessel{\el}(2\pi
\|\blinel\|)$ is non-zero for the particular value of $\el$ and the
radius $\|\blinel\|$ considered.  Firstly, let us attempt to recover
$\shc{\beammodintenl}{\el}{\m}$ for all $\el$ and $\m$ from a
spherical sampling of the visibility function at a single radius
$\|\blinel\|$.  In theory this is possible since the $z$th zeros of
the spherical Bessel function are strictly monotonically increasing
with order $\el$ \citep{liu:2007}, hence by choosing a value of
$2\pi\|\blinel\|$ that lies between two zeros of identical zero-order
$z$ and adjacent function-order $\el$, we ensure that we avoid all
zeros of the spherical Bessel functions.  Nevertheless, as the order
$\el$ deviates from the adjacent values chosen, the spherical Bessel
functions will become arbitrarily closer to zero.  Consequently, any
numerical attempts to recover $\shc{\beammodintenl}{\el}{\m}$ using
this procedure will be unstable.  A full $\blinel$ sampling of the
visibility function in $\reals^3$ is therefore required to recover the
full-sky beam-modulated intensity function.  For each value of $\el$,
sampled spherical surfaces with different radii should be used.  The
value $2\pi \|\blinel\|$ used for a particular $\el$ should be chosen
to ensure that it does not lie on or near the zeros of the spherical
Bessel function $\sbessel{\el}(2\pi \|\blinel\|)$, and ideally at a
low-order local extremum of the spherical Bessel function.

We have demonstrated that it is possible to recover the beam-modulated
intensity over the entire sky in theory, however to do this we require
full sampling of the visibility function in $\reals^3$.  For
interferometric observations, as the field of view rotates over the
sky with time we sample the visibility function for various values of
the baseline in local coordinates.  Typically these samples lie on a
series of $uv$-tracks in the $\blinel=(u,v,w)$ space for values of $w$
close to zero.  For low pointing directions we recover samples at
values of $w$ further from zero, but we are unlikely in practice to
recover sufficient sampling of the visibility function to reconstruct
the beam-modulated intensity function over the entire sky.
Consequently, we do not advocate the use of the spherical harmonic
representations derived above for image recovery.

\subsection{Spherical Haar wavelet space representation}
\label{sec:fsi_shw}

The main challenges that arise from the full-sky interferometry
formulations in both real and spherical harmonic space are related to
the localisation of signal characteristics.  In real space we achieve
good spatial localisation but poor frequency localisation; in
spherical harmonic space we achieve good frequency localisation but
poor spatial localisation.  Both the primary beam and the sky
intensity functions are characterised by spatially localised high
frequency content, thus neither real nor harmonic space bases provide
an efficient representation of these functions.  However, these
functions can be represented efficiently in a wavelet basis due to the
simultaneous spatial and scale localisation afforded by a wavelet
analysis (as discussed in \sectn{\ref{sec:background_wavelets}}).
We derive here the SHW representation of the interferometry integral
and discuss the implications of this representation for computing
visibilities and for image reconstruction.

\subsubsection{Computing visibilities}
\label{sec:fsi_shw_vis}

Consider the SHW representation of the beam-modulated intensity function
\begin{equation}
\beammodintenl(\sal) =
\sum_{\locat=0}^{\npix_{\scalmin} - 1} 
\acoeff_{\scalmin,\locat} \: \scalefun_{\scalmin,\locat}(\sal)
+
\sum_{\scal=\scalmin}^{\scalmax-1}
\sum_{\locat=0}^{\npix_\scal - 1}
\sum_{\wavtype=0}^{2}
\dcoeff_{\scal,\locat}^\wavtype \:
\wav_{\scal,\locat}^\wavtype(\sal)
\spcend ,
\end{equation}
where $\acoeff_{\scalmin,\locat}$ and
$\dcoeff_{\scal,\locat}^\wavtype$ are the scaling and wavelet
coefficients of the beam-modulated intensity function respectively,
and the SHW representation of the plane wave
\begin{equation}
\label{eqn:wave_shw}
\exp{- \img 2 \pi \blinel \cdot \sal} =
\sum_{\locat=0}^{\npix_{\scalmin} - 1} 
\acoeffe_{\scalmin,\locat}(\blinel) \: \scalefun_{\scalmin,\locat}(\sal)
+
\sum_{\scal=\scalmin}^{\scalmax-1}
\sum_{\locat=0}^{\npix_\scal - 1}
\sum_{\wavtype=0}^{2}
\dcoeffe_{\scal,\locat}^\wavtype(\blinel) \:
\wav_{\scal,\locat}^\wavtype(\sal)
\spcend ,
\end{equation}
where $\acoeffe_{\scalmin,\locat}$ and
$\dcoeffe_{\scal,\locat}^\wavtype$ are the scaling and wavelet
coefficients of the plane wave respectively.  Substituting these
expansions into the local coordinate visibility integral given by
\eqn{\ref{eqn:fsi_local}}, we obtain
\begin{align*}
&\vis(\blinel) =
 \sum_{\locat=0}^{\npix_{\scalmin} - 1}
\sum_{\locat\p=0}^{\npix_{\scalmin} - 1}
\acoeff_{\scalmin,\locat} \:
\acoeffe_{\scalmin,\locat\p}(\blinel)
\int_\sphere \scalefun_{\scalmin,\locat}(\sal) \: \scalefun_{\scalmin,\locat\p}(\sal) \dmu{\sal}
\\
&\, + 
\sum_{\locat=0}^{\npix_{\scalmin} - 1}
\sum_{\scal\p=\scalmin}^{\scalmax-1}
\sum_{\locat\p=0}^{\npix_{\scal\p} - 1}
\sum_{\wavtype\p=0}^{2}
\acoeff_{\scalmin,\locat} \:
\dcoeffe_{\scal\p,\locat\p}^{\wavtype\p}(\blinel)
\int_\sphere \scalefun_{\scalmin,\locat}(\sal) \: \wav_{\scal\p,\locat\p}^{\wavtype\p}(\sal) \dmu{\sal}
\\
&\, +
\sum_{\scal=\scalmin}^{\scalmax-1}
\sum_{\locat=0}^{\npix_\scal - 1}
\sum_{\wavtype=0}^{2}
\sum_{\locat\p=0}^{\npix_{\scalmin} - 1}
\dcoeff_{\scal,\locat}^\wavtype \:
\acoeffe_{\scalmin,\locat\p}(\blinel)
\int_\sphere \wav_{\scal,\locat}^{\wavtype}(\sal) \: \scalefun_{\scalmin,\locat\p}(\sal)  \dmu{\sal}
\\
& \, +
\sum_{\scal=\scalmin}^{\scalmax-1}
\sum_{\locat=0}^{\npix_\scal - 1}
\sum_{\wavtype=0}^{2}
\sum_{\scal\p=\scalmin}^{\scalmax-1}
\sum_{\locat\p=0}^{\npix_{\scal\p} - 1}
\sum_{\wavtype\p=0}^{2}
\dcoeff_{\scal,\locat}^\wavtype \:
\dcoeffe_{\scal\p,\locat\p}^{\wavtype\p}(\blinel)
\int_\sphere \wav_{\scal,\locat}^{\wavtype}(\sal) \: \wav_{\scal\p,\locat\p}^{\wavtype\p}(\sal)  \dmu{\sal}
.
\end{align*}
Noting the orthogonality of the scaling functions and wavelets this
reduces to
\begin{equation}
\label{eqn:fsi_shw}
\vis(\blinel) =
\sum_{\locat=0}^{\npix_{\scalmin} - 1}
\acoeff_{\scalmin,\locat} \:
\acoeffe_{\scalmin,\locat}(\blinel)
+
\sum_{\scal=\scalmin}^{\scalmax-1}
\sum_{\locat=0}^{\npix_\scal - 1}
\sum_{\wavtype=0}^{2}
\dcoeff_{\scal,\locat}^\wavtype \:
\dcoeffe_{\scal,\locat}^{\wavtype}(\blinel)
\spcend .
\end{equation}
Applying \eqn{\ref{eqn:fsi_shw}} naively to compute visibilities in
the full-sky setting is no more efficient than computing visibilities
in real or harmonic space and scales as $\order(\npix)$ (since the
multiresolution SHW decomposition contains exactly as many scaling and
wavelet coefficients as the number of pixels on the original sphere).
Furthermore, to compute \eqn{\ref{eqn:fsi_shw}} one must first compute
the wavelet coefficients of the plane wave though
\eqn{\ref{eqn:wave_shw}}, for each baseline $\blinel$.  The spherical
harmonics define a natural harmonic representation on the sphere,
arising from the solution to Laplace's equation in spherical
coordinates.  Consequently, the expansion of the plane wave $\exp{\img
  2\pi \blinel \cdot \sa}$ can be performed analytically in the
spherical harmonic basis through \eqn{\ref{eqn:wave_sh}}.
Unfortunately this is not possible in the SHW basis and the scaling
and wavelet coefficients of the plane wave must be computed
numerically, introducing an additional overhead when computing
visibilities in SHW space.  However, these disadvantages are more than
offset by the efficient representation of the beam-modulated intensity
function in SHW space.  Beam-modulated intensity functions are likely
to contain localised high-frequency content and hence will be
extremely sparse in the wavelet basis, \ie\ a large number of wavelet
coefficients will be zero.  If we consider only the non-zero wavelet
coefficients when computing \eqn{\ref{eqn:fsi_shw}}, visibilities can
be computed at substantially lower computational expense.  Due to the
pairing of approximation and detail coefficients in
\eqn{\ref{eqn:fsi_shw}}, it is also only necessary to compute the
scaling and wavelet coefficients of the plane wave that correspond to
the non-zero coefficients of the beam-modulated intensity function,
thus reducing the overhead discussed above.  Although it is not
possible to determine exactly how \eqn{\ref{eqn:fsi_shw}} scales with
the resolution of the pixelised spheres considered, since this depends
on the SHW sparsity properties of the particular beam-modulated
intensity function considered, the complexity of computing
\eqn{\ref{eqn:fsi_shw}} is found (see \sectn{\ref{sec:sim}}) to
something scale like $\order(\elmax^n)$, where $n \lesssim 1$.

In practice, not only are many of the SHW coefficients of the
beam-modulated intensity identically zero, but many are also very
close to zero.  These wavelet coefficients contain minimal information
content and can be set identically to zero without introducing
substantial errors in the representation of the original
beam-modulated intensity function on the sphere.  For practical
implementations a strategy is required to determine those wavelet
coefficients that are sufficiently close to zero that they can be
safely ignored.  We develop two such strategies.  The first strategy
involves simple hard thresholding, so that all wavelet coefficients
below a threshold (in absolute value) are ignored.  The threshold
value is determined by specifying the proportion of wavelet
coefficients to retain.  Typically less than one percent of the
wavelet coefficients can be kept while still ensuring visibilities are
computed accurately, as we shall see in \sectn{\ref{sec:sim}}.  This
strategy treats all of the wavelet coefficients identically.  However,
wavelet coefficients on coarser levels (lower \scal) are defined over
a larger portion of the sky and so contain more information content.
One should therefore favour keeping wavelet coefficients at coarser
levels over finer levels.  Our second strategy for determining the
wavelet coefficients to retain is based on hard thresholding with an
annealing strategy to determine the threshold value separately for
each level \scal.  The proportion of coefficients to retain at the
finest level is specified and this proportion is increased
quadratically as ones progresses to coarser levels.  We also
experimented with linear and exponential annealing strategies, however
the quadratic strategy was most successful in characterising the
importance of wavelet coefficients across levels.

In \sectn{\ref{sec:sim}} low-resolution simulations of full-sky
interferometric observations are computed using the SHW method
outlined here.  Due to the superior computational efficiency of the
SHW method compared to the real and spherical harmonic space
approaches, it is also possible to perform high-resolution simulations
of interferometric observations using this method; these are also
presented in \sectn{\ref{sec:sim}}.

\subsubsection{Image reconstruction}
\label{sec:fsi_shw_recon}

In the SHW formulation of full-sky interferometry given by
\eqn{\ref{eqn:fsi_shw}}, the scaling and wavelet coefficients of the
beam-modulated intensity function are linearly related to the
visibilities by the scaling and wavelet coefficients of the plane
wave.  The plane wave is oscillatory and does not contain localised
high frequency content, hence the plane wave is not likely to be
sparsely represented in the SHW basis.  Consequently, the use of
\eqn{\ref{eqn:fsi_shw}} for wide field of view image reconstruction is
much more well posed than the spherical harmonic representation of the
visibility integral.  Furthermore, if the primary beam is known, which
is often the case, then one need only attempt to recover the wavelet
coefficients of the beam-modulated intensity function that correspond
to the non-zero coefficients of the primary beam, \ie\ the beam tells
us exactly which wavelet coefficients to recover and all others can
safely be assumed to be zero.

SHW image reconstruction on the full-sky therefore involves inverting the
linear system given by \eqn{\ref{eqn:fsi_shw}}.  To solve this system
is it instructive to write it as the matrix equation
\begin{equation*}
\vis(\blinel) =
\begin{bmatrix}
\bmath{\acoeffe}_{\scalmin}(\blinel) \\
\bmath{\dcoeffe}(\blinel)
\end{bmatrix}^{\rm T}
\begin{bmatrix}
\bmath{\acoeff}_{\scalmin} \\
\bmath{\dcoeff}
\end{bmatrix}
=
\Delta^{\rm T}(\blinel) \: \bmath{\Gamma}
\spcend ,
\end{equation*}
where $\bmath{\acoeff}_{\scalmin}$ and $\bmath{\dcoeff}$ are vectors
of concatenated non-zero scaling and wavelet coefficients of the
beam-modulated intensity and $\bmath{\acoeffe}_{\scalmin}(\blinel)$
and $\bmath{\dcoeffe}(\blinel)$ are the corresponding coefficients of
the plane wave.  Combining the scaling and detail coefficients into a
single vector we obtain 
$\bmath{\Gamma} = [\bmath{\acoeff}_{\scalmin} \,\, \bmath{\dcoeff}]^{\rm T}$
and
$\bmath{\Delta}(\blinel) = [\bmath{\acoeffe}_{\scalmin}(\blinel) \,\,
\bmath{\dcoeffe}(\blinel)]^{\rm T}$.  Now assuming that we have $M$
visibility observations corresponding to different baselines, we may
write the system of equations that we recover as the single matrix equation
\begin{equation}
\label{eqn:fsi_shw_matrix}
\bmath{\vis} =
\mathbf{M}^{\rm T} \: \bmath{\Gamma}
\spcend ,
\end{equation}
where
\begin{equation*}
\bmath{\vis} =
\begin{bmatrix}
\vis(\blinel_0) & 
\vis(\blinel_1) &
\cdots & 
\vis(\blinel_M)
\end{bmatrix}^{\rm T}
\end{equation*}
and
\begin{equation*}
\mathbf{M} = 
\begin{bmatrix}
\bmath{\Delta}(\blinel_0) & 
\bmath{\Delta}(\blinel_1) &
\cdots & 
\bmath{\Delta}(\blinel_M)
\end{bmatrix}
\spcend .
\end{equation*}
The overdetermined system \eqn{\ref{eqn:fsi_shw_matrix}}
may solved in the least squares sense for an estimate of the
non-zero scaling and wavelet coefficients of the beam-modulated
intensity function:
\begin{equation}
\bmath{\Gamma}_{\rm LS} = \bigl(\mathbf{M} \: \mathbf{M}^{\rm T}\bigr)^{-1} \:
\mathbf{M} \: \bmath{\vis}
\spcend .
\end{equation}

One would expect that recovering the beam-modulated intensity function
on the full-sky in this manner would be well posed, however to really
ascertain the effectiveness of the method outlined here it should be
implemented and tested.  The focus of the current article is
predominantly on the forward interferometric wide field imaging
problem.  The use of SHWs for wide field of view image reconstruction
is an interesting application in its own right and we intend to
develop these ideas further and present the results of various
experiments in a future work.

\subsection{Incorporating horizon occlusion and variable beams}
\label{sec:fsi_horizon}

In the preceding formulations we have assumed that the full-sky is
visible.  Obviously this is not the case for Earth based
interferometers that may observe only the hemisphere above the
horizon.  Nevertheless, if the interferometer pointing direction is
relatively close to the North pole of the observable hemisphere and
the beam is sufficiently small that it is zero in the southern
hemisphere, then the full-sky formulation is appropriate.
Furthermore, we have assumed that the primary beam is fully defined in
local coordinates and is independent of the interferometer pointing
direction on the sky.  This is unlikely to be the case for real
aperture array interferometers.  In this section we formulate full-sky
interferometry in the setting where the beam is sufficiently large
that it may be non-zero below the horizon and we discuss extensions to
incorporate primary beams that also depend on pointing direction.

In order to incorporate horizon occlusion and variable beams we must
introduce a third coordinate system.  In addition to the local
coordinate system defined relative to the interferometer pointing
direction and the global coordinate system of the celestial sky, it is
necessary to introduce an Earth-based coordinate system.  We define
the Earth-based coordinate frame by the right handed set of unit
vectors $\{\eone,\etwo,\ethree\}$.  The Earth-based coordinate system
is related to the celestial sky coordinate system by a time varying
rotation $\rott$ that relates the orientation of the celestial sky
relative to the Earth.  Let $\rotmatt$ denote the corresponding time
varying $3\times3$ rotation matrix relating these coordinate systems.
The unit vectors defining these coordinate systems are then related by
$\nind=\rotmatt\eind$.  We adopt the convention that these coordinate
frames are aligned for time $t=0$, \ie\ $\rotmatt{}_{=0}=\mathbf{I}$,
where $\mathbf{I}$ is the identity matrix.  A vector in Earth-based
coordinates is related to a vector in celestial sky coordinates by
$\sae=\rotmatt^{-1} \san$, where the superscript $\erth$ denotes
Earth-based coordinates.  In particular, the interferometer pointing
direction in Earth-based coordinates traces a fixed point in the
celestial sky $\saon$ over time: $\saoe(t)=\rotmatt^{-1} \saon$, \ie\
the pointing direction in Earth-based coordinates is time dependent.
The local coordinate system is now defined relative to $\saoe(t)$ and
is related to the Earth-based coordinate system by the rotation
$\roto\equiv \rot(\sab^\erth,\saa^\erth,0)$, with corresponding
rotation matrix $\rotmato$, where $(\saa^\erth,\sab^\erth)$ are the
spherical coordinates of $\saoe(t)$.  For notational simplicity the
time dependence of $\roto$ and $\rotmato$ is left implicit.

Horizon occlusion may be modelled by incorporating a binary horizon
function that is unity above the Earth's horizon and zero below.  This horizon
function is represented most naturally in the Earth-based coordinate system
$\horizone(\sae)\in\ltwo(\sphere,\dmun)$ and is defined by
\begin{align*}
\horizone(\sae) = 
\begin{cases}
 1 &  \text{if} \;\; \sae\cdot\ethree>0 \\
 0 &  \text{otherwise .}
\end{cases}
\end{align*}
Representing each function in its natural coordinate system the
interferometer visibility integral may be written
\begin{equation}
\label{eqn:vis_horizon}
\vis(\blinel) = \int_\sphere 
\beaml(\sal) 
\horizone(\sae)
\intenn(\san)
\exp{- \img 2 \pi \blinel \cdot \sal}
\dmu{\sal}
\spcend ,
\end{equation}
where the binary horizon function is included to exclude contributions
to the visibility integral from below the horizon.  It is again
convenient to represent all functions in the visibility integral in a
consistent coordinate system.  In local coordinates the horizon
function becomes
\begin{equation*}
\horizone(\sae) = 
\horizone(\rotmato \sal) = 
(\roto^{-1} \horizone)(\sal) = 
\horizonl(\sal)
\spcend ,
\end{equation*}
\ie\ $\horizonl = \roto^{-1} \horizone$, and the source intensity function becomes
\begin{equation*}
\intenn(\san) = 
\intenn(\rotmatt \rotmato \sal) = 
(\roto^{-1} \rott^{-1} \intenn)(\sal) = 
\intenl(\sal)
\spcend ,
\end{equation*}
\ie\ $\intenl = \roto^{-1} \rott^{-1} \intenn$.  
Any azimuthal $\eulc$ rotation of $\rott$ will render $\intenl$ time
dependent due to the fixed azimuthal rotational component of $\roto$.

The local coordinate version of the visibility integral then follows
trivially from \eqn{\ref{eqn:vis_horizon}}.  The techniques outlined
in \sectn{\ref{sec:fsi_coord}} through \sectn{\ref{sec:fsi_shw}} can
then be applied directly by replacing the beam-modulated intensity
function $\beammodintenl(\sal)$ with the horizon-beam-modulated
intensity function $\beamhorzmodintenl(\sal)=\beaml(\sal)
\horizonl(\sal) \intenl(\sal)$.
%
The time dependence of $\horizonl$ and $\intenl$ now precludes
any full-sky reconstruction of the beam-horizon-modulated intensity
function even in theory, let alone in practice.  The interferometer
baseline in local coordinates is given by
$\blinel=\rotmato^{-1}\blinee$, where $\blinee$ is the natural
representation of the baseline in Earth-based coordinates.  Note that
$\blinel$ is time dependent due to the time dependence of $\rotmato$.
It was not possible to define $\blinel$ explicitly previously since we
did not have a natural coordinate system to represent the
interferometer baseline (due to the absence of the Earth-based
coordinate system).

If the primary beam function depends on the pointing direction in
Earth-based coordinates then it also becomes time dependent, \ie\
$\beaml(\sal,\saoe(t))$.  In this case the representations of the
visibility integral derived previously in real, spherical harmonic and
SHW space all still hold, however $\beaml(\sal)$ must be recomputed
for each pointing direction.  In this framework it is also possible to
include more complicated horizon effects in the primary beam, rather
than the simple masking adopted by including the binary horizon
function.  A simulated interferometer visibility observation
incorporating horizon occlusion and a variable beam would be performed
as follows:

\begin{enumerate}

\item compute the Earth-based pointing direction by
 \mbox{$\saoe(t)=\rotmatt^{-1} \saon$}; 

\item compute the interferometer baseline in local coordinates by
  $\blinel=\rotmato^{-1}\blinee$;

\item compute the intensity function in local coordinates by
  \mbox{$\intenl = \roto^{-1} \rott^{-1} \intenn$};

\item compute the horizon function in local coordinates by 
  \mbox{$\horizonl = \roto^{-1} \horizone$};

\item compute the primary beam function for the given pointing
  direction by $\beaml(\sal,\saoe(t))$;

\item compute the observed visibility $\vis(\blinel)$ by the method of
  choice using either \eqn{\ref{eqn:fsi_local}}, \eqn{\ref{eqn:fsi_harm}}
  or \eqn{\ref{eqn:fsi_shw}};

\item repeat steps (i)-(vi) for each observation time $t$.

\end{enumerate}

Although the beam may be sufficiently large that a small patch
approximation is not valid, it may often be the case that it is not so
large that horizon occlusion must be included when computing
visibilities.  Consequently, although we have presented the most
general formulation of full-sky interferometry in this section by
carefully modelling horizon occlusion and including variable beams, it
may often be appropriate to neglect these effects and simply follow
the formulation outlined previously.


\section{Simulations}
\label{sec:sim}

We perform simulations of interferometric observations, including
full-sky contributions, in order to validate the full-sky
interferometry formulations presented in \sectn{\ref{sec:fsi}}.
Firstly, we compare all of the methods on low-resolution simulations.
Performing full-sky interferometric simulations using either the real
or spherical harmonic space representation is a substantial
computational challenge and is not feasible for high-resolution
simulations.  The SHW methods ease this computational burden
considerably, rendering higher resolution simulations feasible.  Using
SHW methods we also perform high-resolution simulations of full-sky
interferometric observations.  Before presenting these low- and
high-resolution simulations we begin with a brief discussion of the
practical considerations that must be taken into account.

\subsection{Practical considerations}
\label{sec:sim_practical}

In this article we focus predominantly on the forward wide field
imaging problem that involves simulating the visibilities observed by
an interferometer when including full-sky contributions.  In terms of
the inverse problem of image reconstruction on wide fields of view, we
have seen that this problem is not well posed in the spherical
harmonic representation but is likely to be well posed in the SHW
space representation.  We have outlined a preliminary method to
perform image reconstruction using SHWs in
\sectn{\ref{sec:fsi_shw_recon}}.  In a future work we intend to
develop these ideas further and to examine the performance of these
methods on simulations.  However, we restrict our attention in the
simulations performed in this article to the forward problem of
simulating visibilities.  Consequently, although we consider full-sky
effects here when computing visibilities, we adopt the usual Fourier
transform approach for recovering images on the tangent plane at the
interferometer pointing direction.  We next relate the parameters of
our visibility simulations, including the visibilities that must be
computed, to the resolution and size of the recovered tangent plane
image.

In practice we assume that all signals on the sphere are band-limited
at $\elmax$.  Although the exact number of samples on the sphere
$\npix$ required to represent a band-limited signal depends on the
pixelisation of the sphere, for the \healpix\ scheme and in general it
is of order $\order(\npix) \sim \order(\elmax^2)$ (\eg\
\citealt{driscoll:1994,gorski:2005}).
The maximum interferometer baseline distance $\blinemax$ for which the
visibility may be computed accurately is related to the harmonic
band-limit \elmax.  For small fields, 
the approximate relationship
$\elmax\simeq2\pi\blinemax$ holds (the exact relationship is
$2\blinemax=\cot(\pi/\elmax)$; \citealt{hobson:1996}).  In order to
ensure Nyquist sampling is satisfied when reconstructing images on
small patches, for square images one requires that $\lx_{\rm
  max}=\sqrt{\nimage}/(2 u_{\rm max})$, where $\lx_{\rm max}$ is the
width of the image plane, $u_{\rm max} = \blinemax / \sqrt{2} $ and
$\nimage$ is the number of pixels of the reconstructed image (these
results generalise to rectangular images trivially).

In order to image these small patches accurately, a high band-limit
$\elmax$ must be considered.  This constraint increases the
computational requirements of computing visibilities in harmonic space
significantly.  
For example, to image an area on the sky of one square degree
($\lx_{\rm max}=1^\circ$) for a 50$\times$50 pixel image
($\nimage=2,500$), we must consider all harmonic coefficients up to a
band-limit of $\elmax\simeq13,000$.  It is then necessary to repeat
this computation for numerous local coordinate representations of the
interferometer baseline $\blinel$ as the source traverses the
sky.
Although an exact sampling theorem does not exist on a \healpix\
sphere, typically the harmonic band-limit of a function sampled on a
\healpix\ sphere is related to the resolution of the sphere through
$\elmax =c \nside$, where $c$ is typically two or three.  In order to
accurately image these small patches in real space while including
full-sky contributions a high resolution pixelisation of the sphere is
therefore required.  For the imaging example considered above, one
must consider all pixel values of a \healpix\ sphere at resolution
$\nside\gtrsim512$. 
As discussed previously, the advantage of the SHW representation is
its simultaneous localisation of signal content in scale and position.
Computing full-sky visibilities using the SHW method adapts to the
sparse representation of the beam-modulated intensity function in the
SHW basis, thus providing substantial computational savings over the
real and spherical harmonic space methods.
We next apply all of these methods to low-resolution simulations to
compare their performance.  Finally, let us note that the visibility
computation for each baseline $\blinel$ is independent and
implementations of all of these methods may be parallelised easily to
spread the computational load.

\subsection{Low-resolution simulated synchrotron observations}
\label{sec:sim_sync}

In this section we simulate low-resolution observations of synchrotron
emission made by an idealised interferometer.  We simulate
visibilities using all of the the full-sky interferometry frameworks
outlined in \sectn{\ref{sec:fsi}}, ensuring that contributions due to
a large primary beam are included.  For simplicity, we do not
incorporate in these simulations the extensions described in
\sectn{\ref{sec:fsi_horizon}} that allow horizon occlusion and
variable beams.  The motivation of these simulations is to demonstrate
and validate the application of the full-sky interferometry
formulations presented previously and to compare the performance of
the methods.  Our implementation of these methods have yet to be
parallelised and all timing tests presented here and subsequently are
performed on a laptop with a single 2.2GHz Intel Core 2 Duo processor
and 2GB of memory.

\begin{figure}
\centering
\subfigure[Mollweide projection]{\includegraphics[width=70mm]{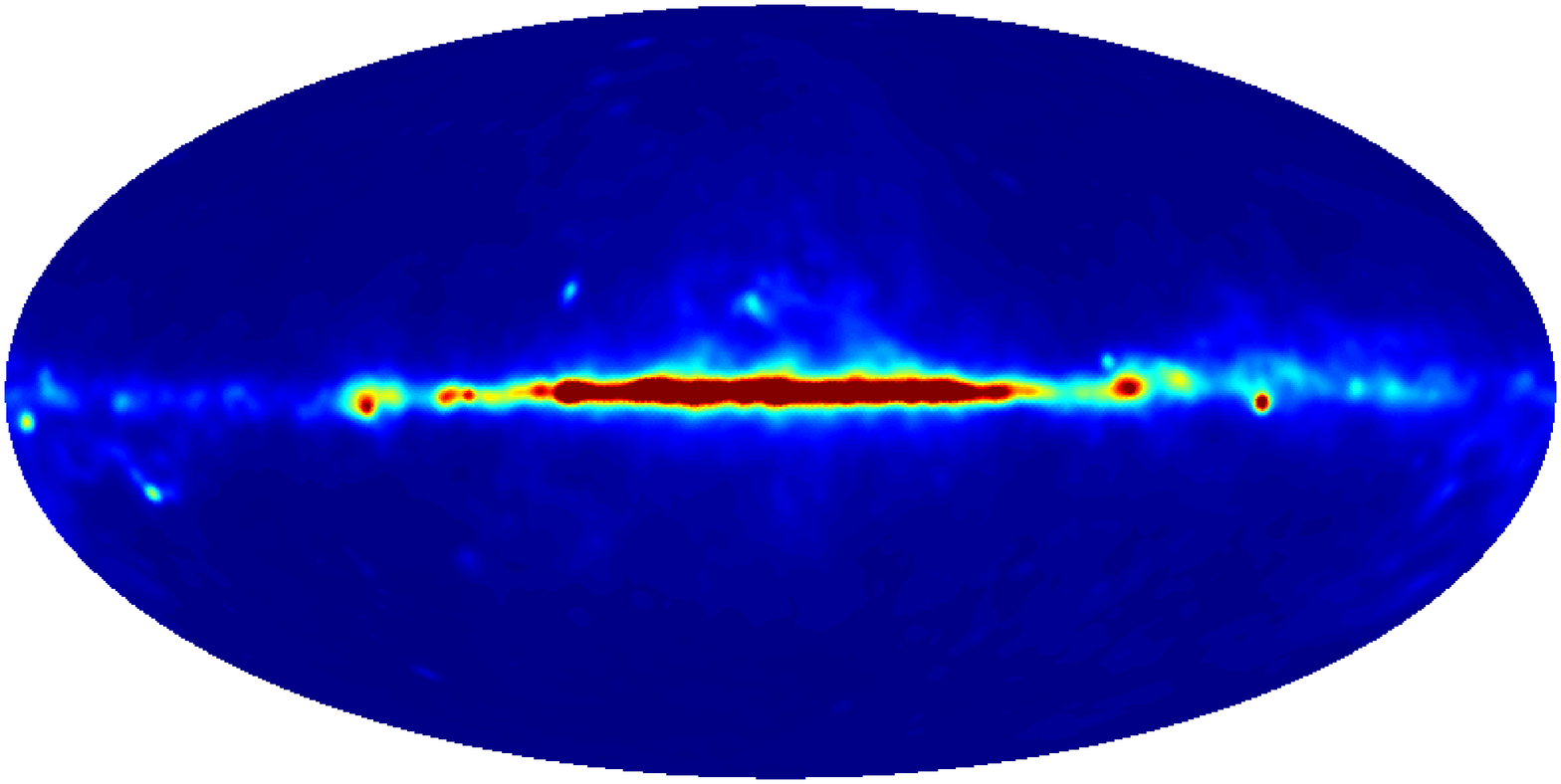}}
\subfigure[Globe]{\includegraphics[clip=,viewport=8 8 640 640,width=50mm]{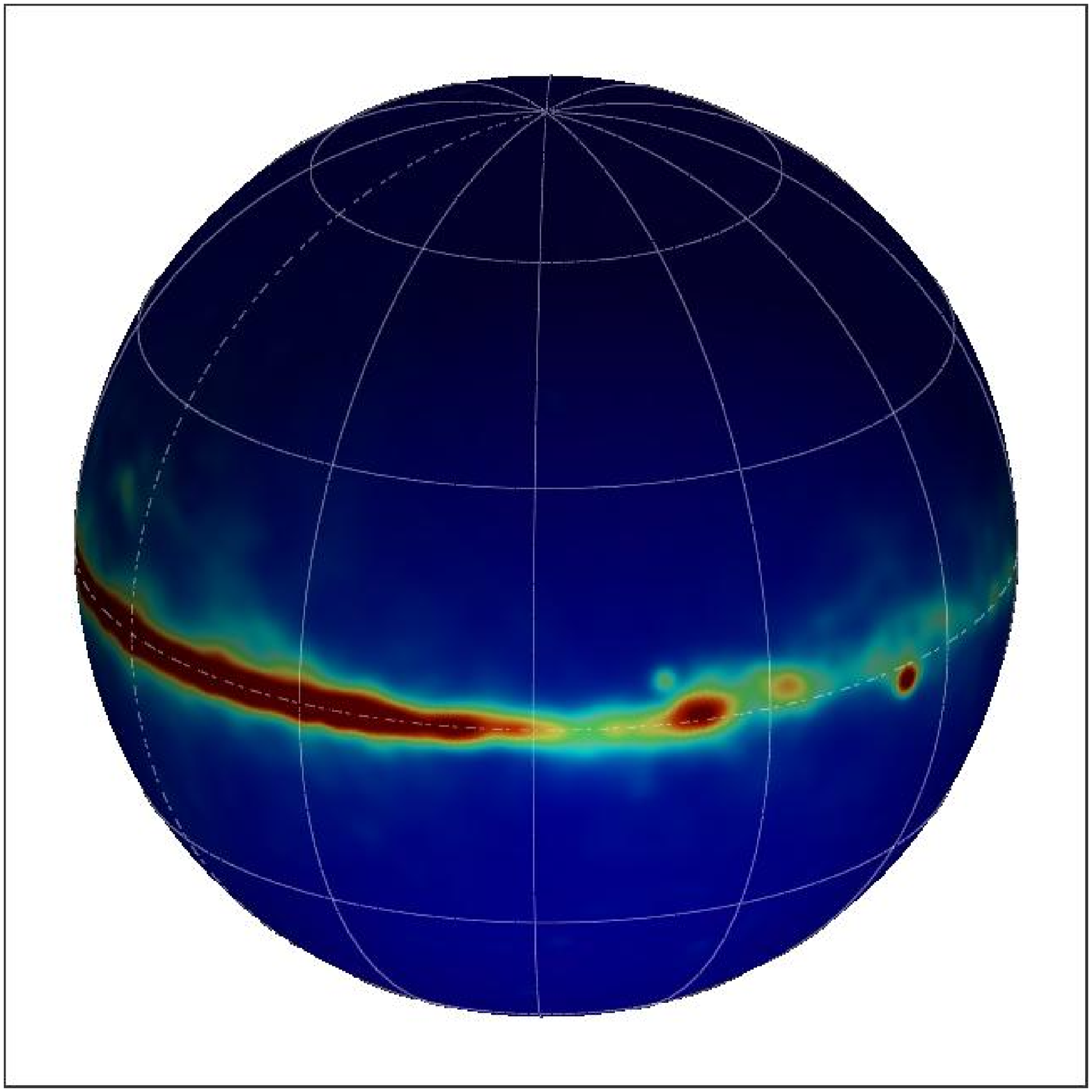}}
\caption{Full-sky synchrotron map observed by \wmap\ and smoothed with
  a Gaussian kernel of ${\rm \fwhm_s}=1.7^\circ$.  This synchrotron
  map provides the full-sky background intensity map for our
  low-resolution simulated observations and is shown here in the
  global coordinate frame defined by Galactic coordinates.}
\label{fig:intensity}
\end{figure}

\begin{figure}
\centering
\subfigure[Synchrotron map]{
\includegraphics[clip=,viewport=8 8 640
640,width=40mm]{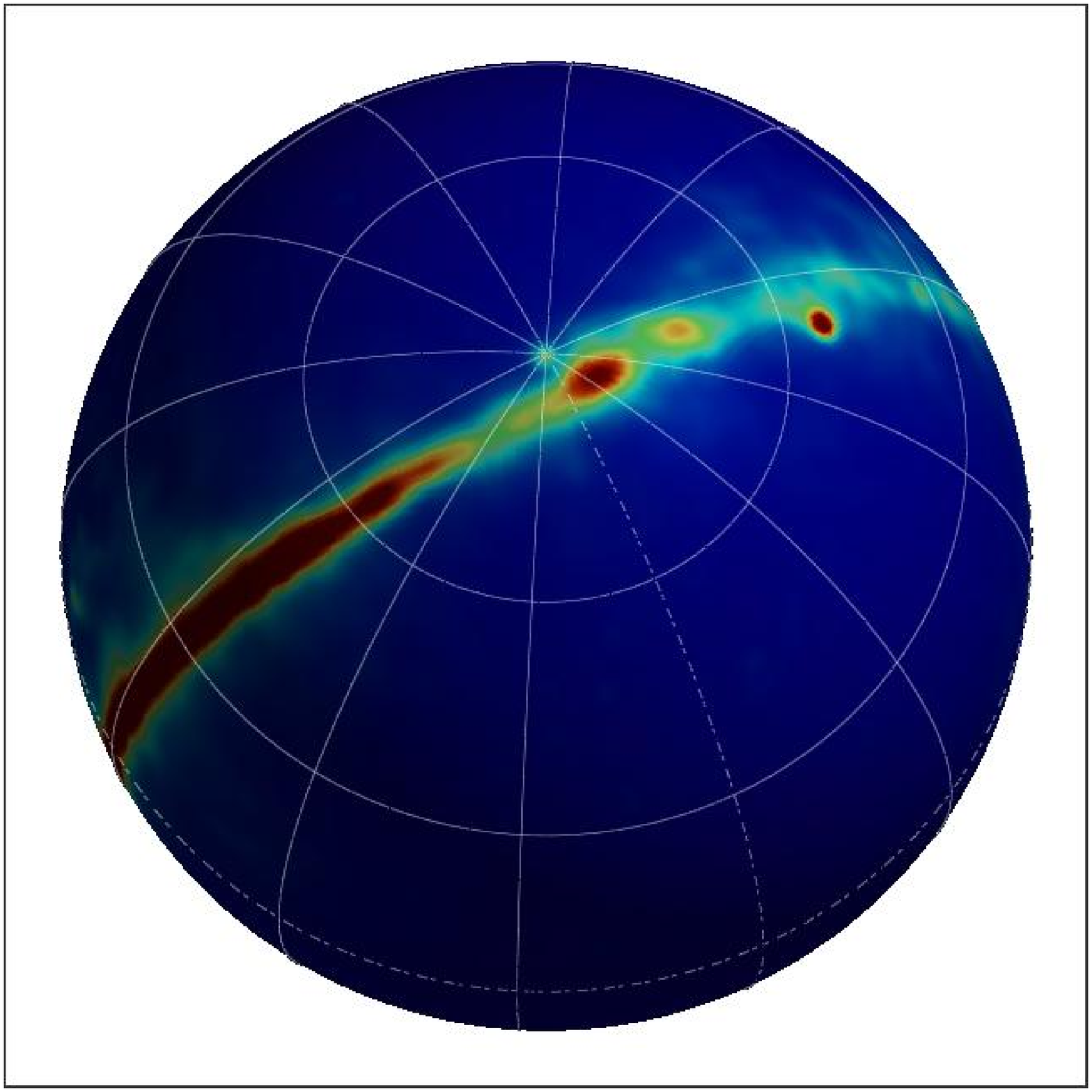} \quad
\includegraphics[width=35mm]{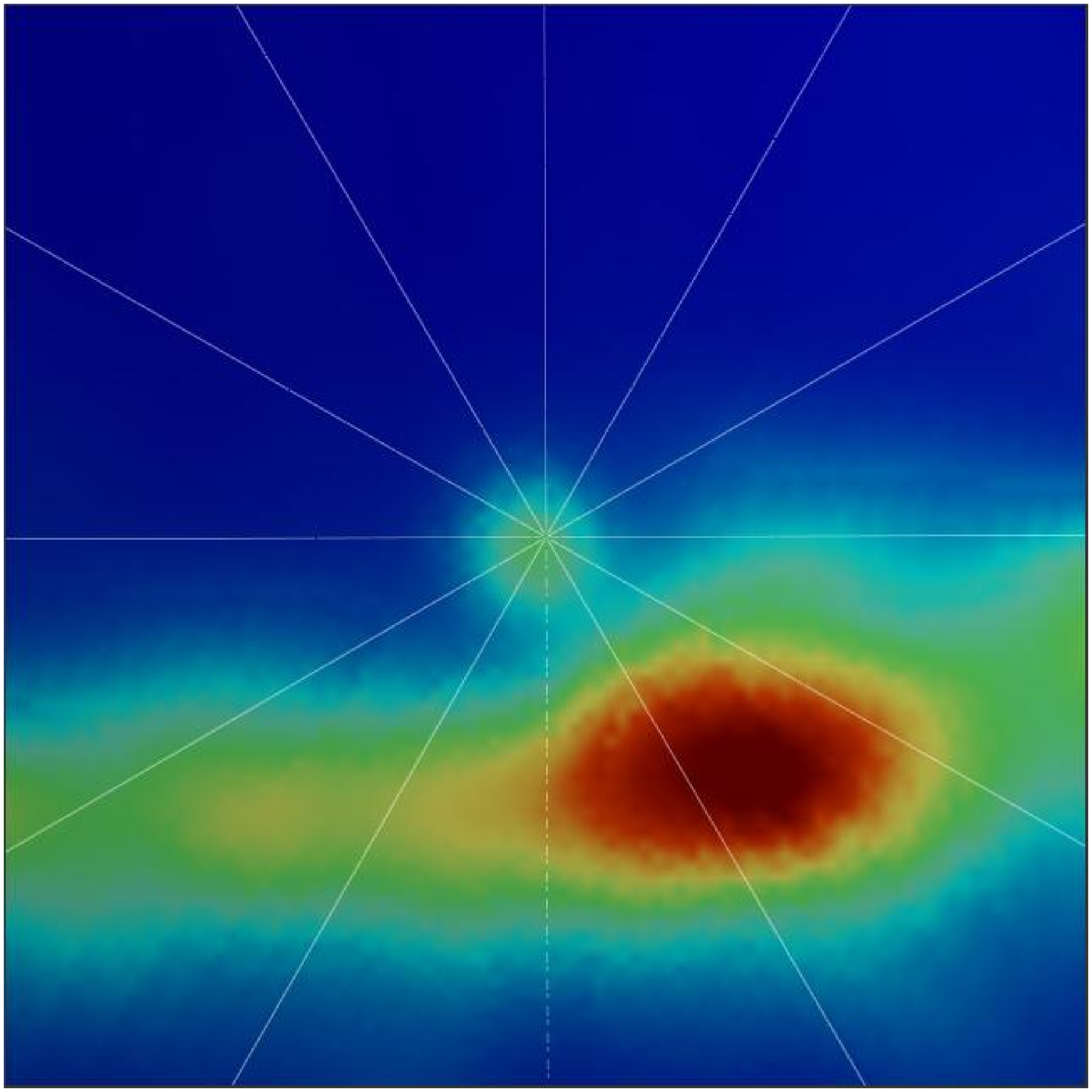}
}
\subfigure[Gaussian beam profile with ${\rm FWHM_b}\simeq18^\circ$]{
\includegraphics[clip=,viewport=8 8 640
640,width=40mm]{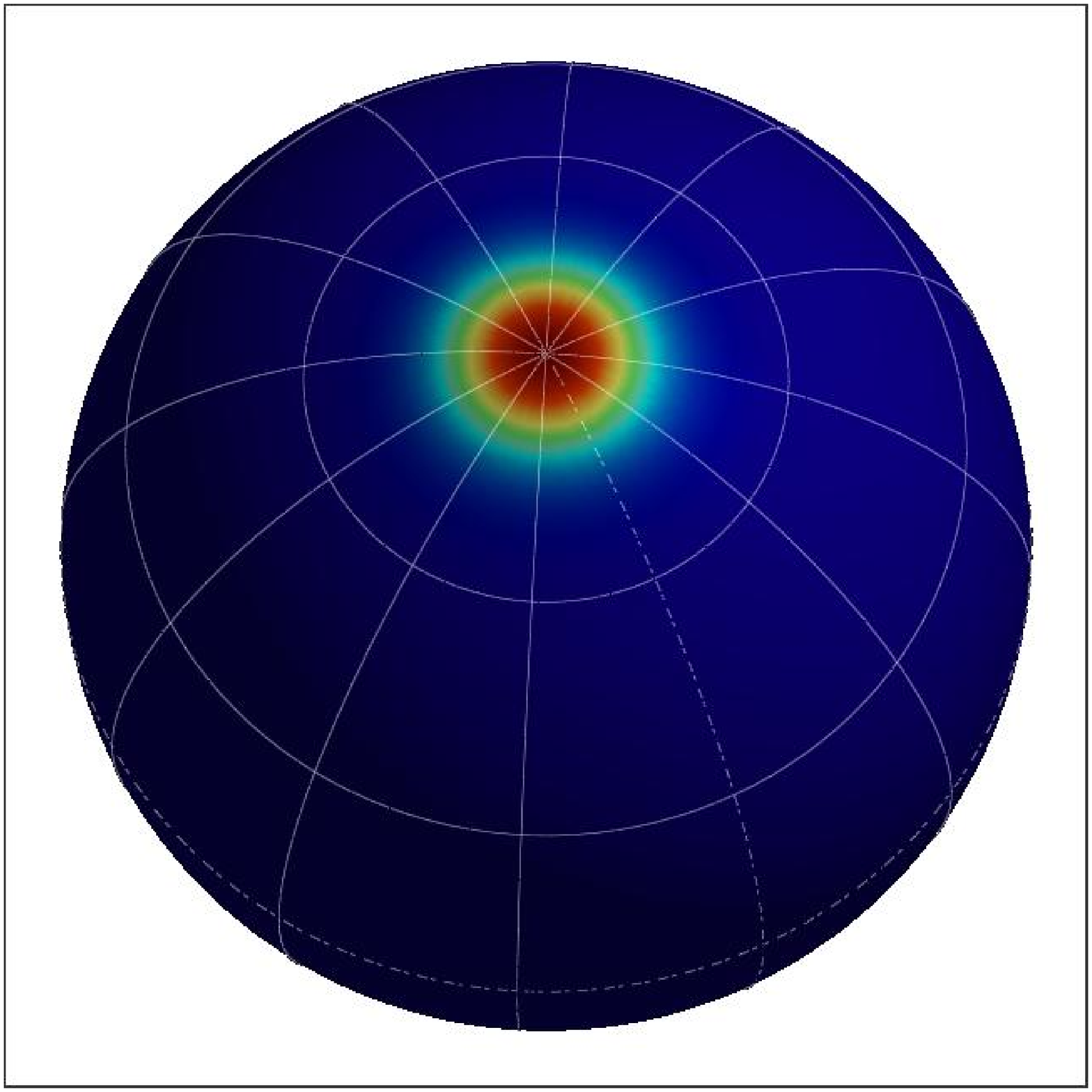} \quad
\includegraphics[width=35mm]{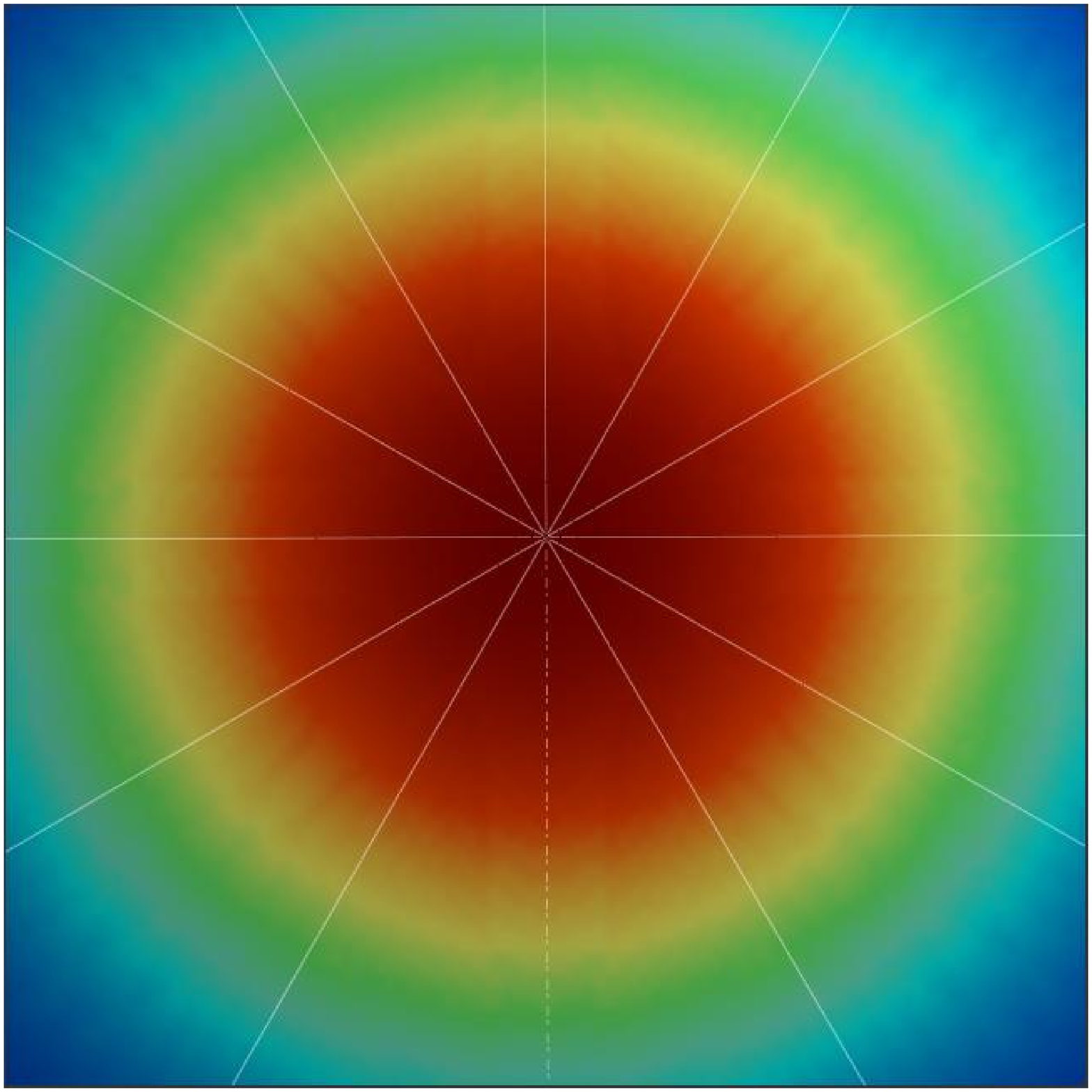}
}
\caption{Full-sky synchrotron and beam maps in local coordinates.  We
  consider the source at $\saon=(\saa_0^\glob,\sab_0^\glob)=(84.0^\circ,76.5^\circ)$.
  The synchrotron map in global coordinates is rotated by $\roto^{-1}\equiv
  \rot(0,-\saa_0^\glob,-\sab_0^\glob)$ to convert to local coordinates.
  Full globes are shown on the left; zoomed images of globes about the north
  pole are shown on the right.  Note that the beam profile takes the
  value of unity in the centre and $\sim\!\!0.3$ at the boundary of the
  zoomed image.
}
\label{fig:north}
\end{figure}

For the background intensity map we use the full-sky synchrotron
template recovered from the 3-year \wmaptext\ (\wmap) observations
\citep{hinshaw:2006}.  A detailed description of the construction of
this template is given by \citet{hinshaw:2006}, however for our
purpose we merely consider this as the full-sky background intensity
map $\inteng(\sag)$ to be observed by our idealised interferometer.
This synchrotron map is available for download from the \lambdaarchtext\
(\lambdaarch).\footnote{\url{http://lambda.gsfc.nasa.gov/}}
In order to remove high-frequency content of this map, we smooth the
data with a Gaussian kernel with full-width-half-maximum of ${\rm
  \fwhm_s}=1.7^\circ$ (since we only perform low-resolution
simulated observations of these data).  The resulting smoothed,
full-sky synchrotron map is illustrated in \fig{\ref{fig:intensity}}.
We simulate observations of the extended source located at
$\saon=(\saa_0^\glob,\sab_0^\glob)=(84.0^\circ,76.5^\circ)$.  In order
to compute full-sky visibility contributions, it is necessary to convert the global
intensity function to its local coordinate version centred on the
interferometer pointing direction.  This is performed through a
rotation by $\roto^{-1}$ as outlined in \sectn{\ref{sec:fsi_coord}}.
The local coordinate intensity function $\intenl(\sal)$ is illustrated
in \fig{\ref{fig:north}~(a)}.  The extended source to be observed is
clearly visible at the north pole of this map.  The interferometer
beam used in these simulations is given by a wide Gaussian with ${\rm
  FWHM_b}\simeq18^\circ$ and is illustrated in
\fig{\ref{fig:north}~(b)}.

Given the mock data and configuration discussed previously, we
simulate visibilities through direct quadrature on the sphere given by
\eqn{\ref{eqn:fsi_local}}, the spherical harmonic space representation
given by \eqn{\ref{eqn:fsi_harm}} and the SHW space representation
given by \eqn{\ref{eqn:fsi_shw}}.  We assume complete $uv$ coverage
and consider a baseline limit of $u_{\rm max} = 30$, corresponding to
$\elmax \simeq 270$, and reconstruct an image of size $\nimage =
20\times20$.  The smoothed synchrotron map is represented by a \healpix\
sampled sphere at resolution $\nside=256$ and so is sampled
sufficiently for these simulations.  Once the visibilities are
simulated, a synthesised image is reconstructed simply by taking the
inverse Fourier transform.  Nyquist sampling dictates that the
synthesised image corresponds to a $\sim20^\circ$ square patch.
This image reconstruction relies on a tangent plane approximation
which will not be valid for the large field of view considered.
Nevertheless, this simple approach to reconstruction is sufficient to
demonstrate the validity of our full-sky interferometry framework in
this restricted low-resolution setting.  The application of the
standard Fourier transform here to synthesise an image from the
simulated visibilities is performed for visual verification only.

The direct projection of the full-sky beam-modulated intensity
function onto the tangent plane at the interferometer pointing
direction
is illustrated in
\fig{\ref{fig:recon_synchrotron}~(a)}.  The images reconstructed from
full-sky visibilities computed using all of the methods outlined in
\sectn{\ref{sec:fsi}} are illustrated in the remaining panels of
\fig{\ref{fig:recon_synchrotron}} (upsampled to a $60\times60$ pixel image
for visualisation).
One would expect the tangent plane image of
\fig{\ref{fig:recon_synchrotron}~(a)} to differ slightly from the
reconstructed images since full-sky contributions due to the large
beam are incorporated when simulating visibilities, however a
flat-patch approximation is assumed when synthesising the image using
the standard Fourier approach.  Furthermore, the images synthesised
from simulated visibilities contain less high-frequency content as a
consequence of the band-limit imposed by the interferometer baseline
limit.  Nevertheless, all images are in close agreement, demonstrating
and validating the use of the methods derived in \sectn{\ref{sec:fsi}}
to simulate visibilities observed by an interferometer in the full-sky
setting.
Furthermore, the images contained in panels (b), (c) and (d) of
\fig{\ref{fig:recon_synchrotron}}, corresponding respectively to the
real, spherical harmonic and naive SHW methods of computing full-sky
visibilities are identical (to numerical precision).  These images are
computed using independent implementations, thereby providing an
additional verification of these methods and implementations.  In
order to realise the advantages of the efficient representation of the
SHW basis, it is necessary to ignore near-zero valued wavelet
coefficients when computing \eqn{\ref{eqn:fsi_shw}}.  This is not done
in the naive SHW method but strategies to disregard unimportant
wavelet coefficients were described in
\sectn{\ref{sec:fsi_shw_recon}}: constant and annealing based
thresholding strategies were proposed.  Reconstructed images based on
these methods are illustrated in panels (e) and (f) of
\fig{\ref{fig:recon_synchrotron}}.  These approaches do introduce a
small error compared to the images of panels (b) through (d) of
\fig{\ref{fig:recon_synchrotron}}, but the controlled introduction of
these errors provide substantial computational savings.

\newlength{\recovplotwidth}
\setlength{\recovplotwidth}{50mm}

\begin{figure*}
\centering
\mbox{
\subfigure[Tangent plane image]{\includegraphics[width=\recovplotwidth]{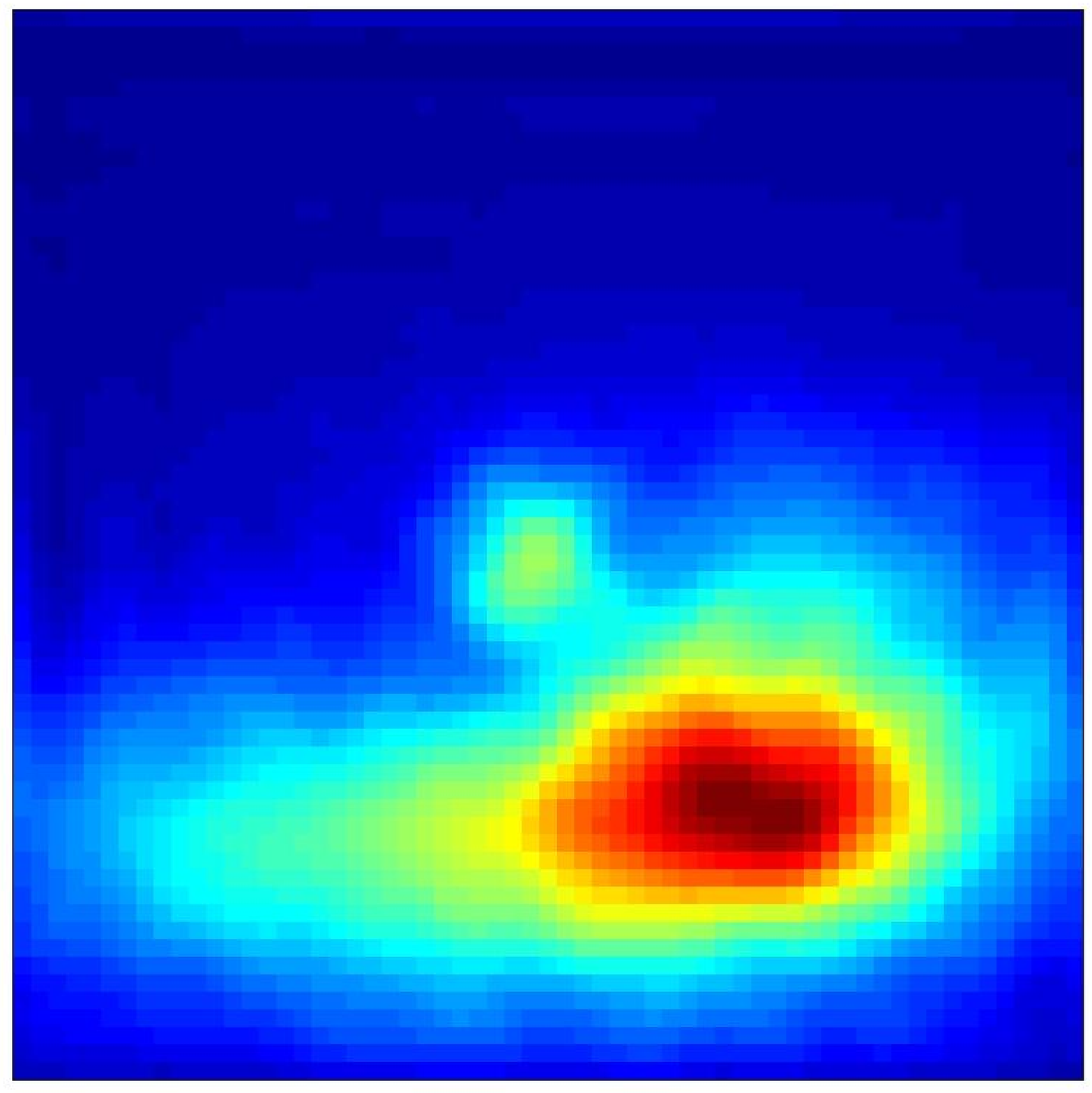}}
\quad
\subfigure[Direct quadrature]{\includegraphics[clip=,viewport=239 83 993 837,width=\recovplotwidth]{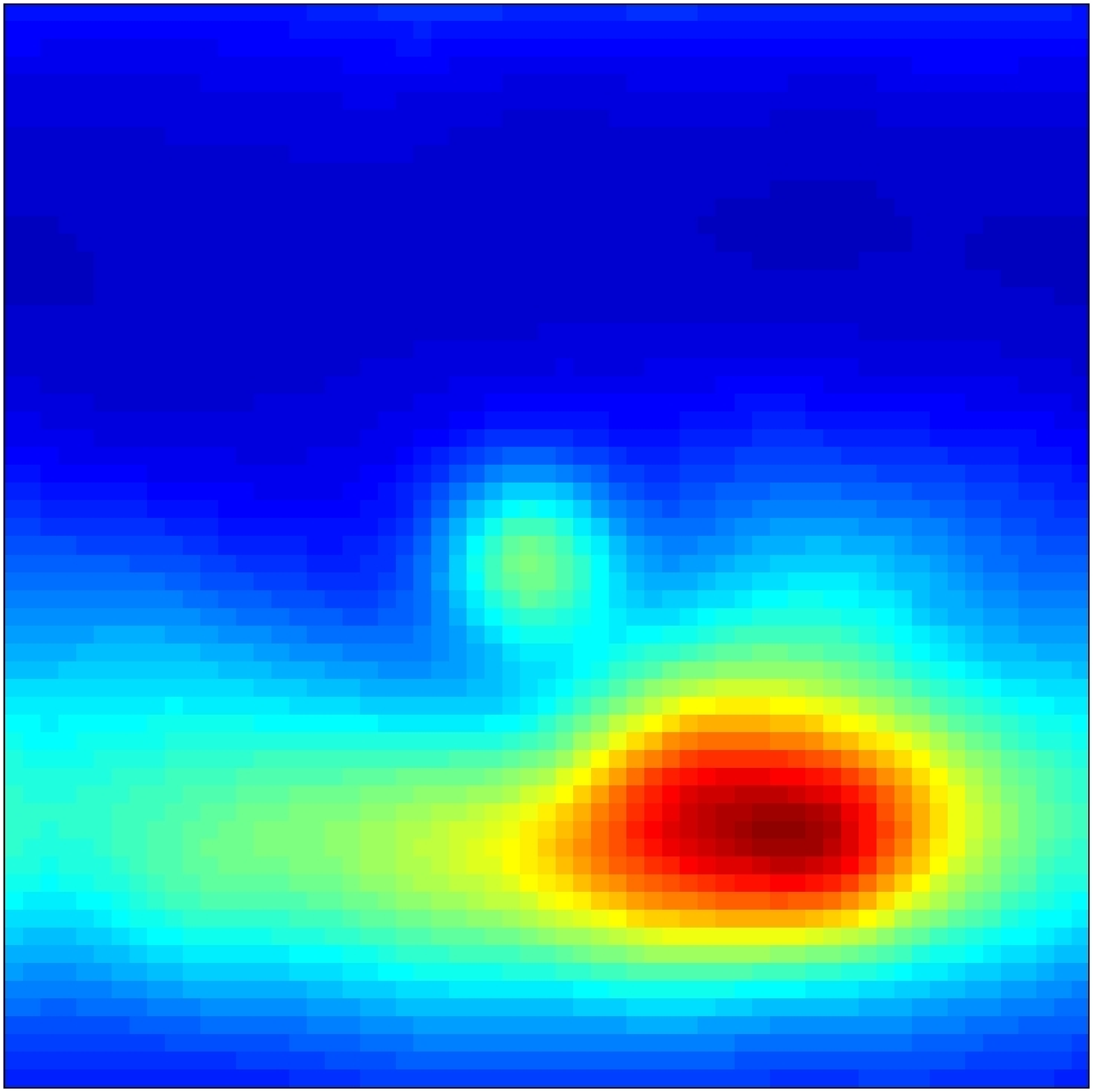}} 
\quad
\subfigure[Spherical harmonics]{\includegraphics[clip=,viewport=239 83 993 837,width=\recovplotwidth]{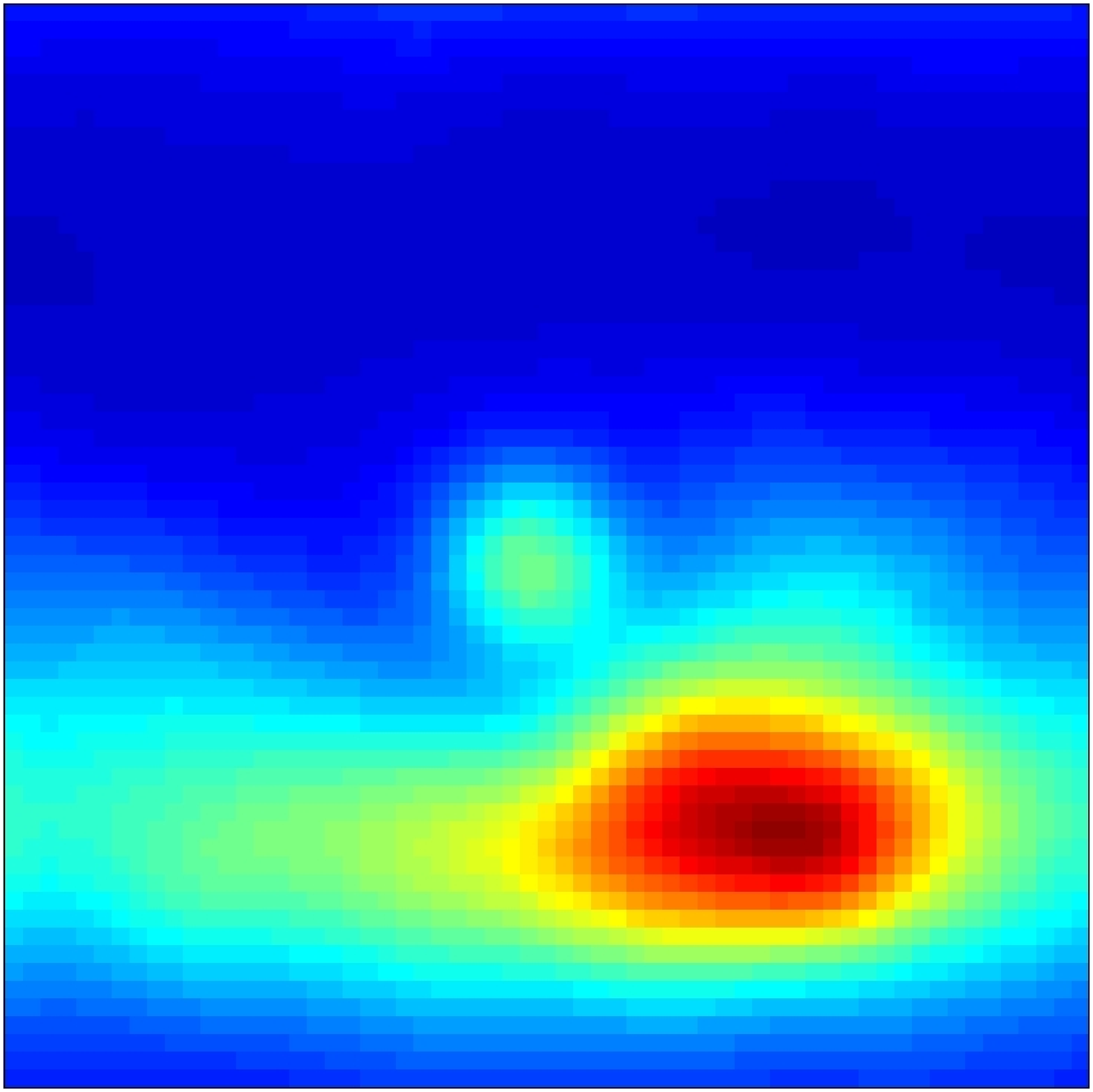}} 
}\\
\mbox{
\subfigure[Naive SHW]{\includegraphics[clip=,viewport=239 83 993 837,width=\recovplotwidth]{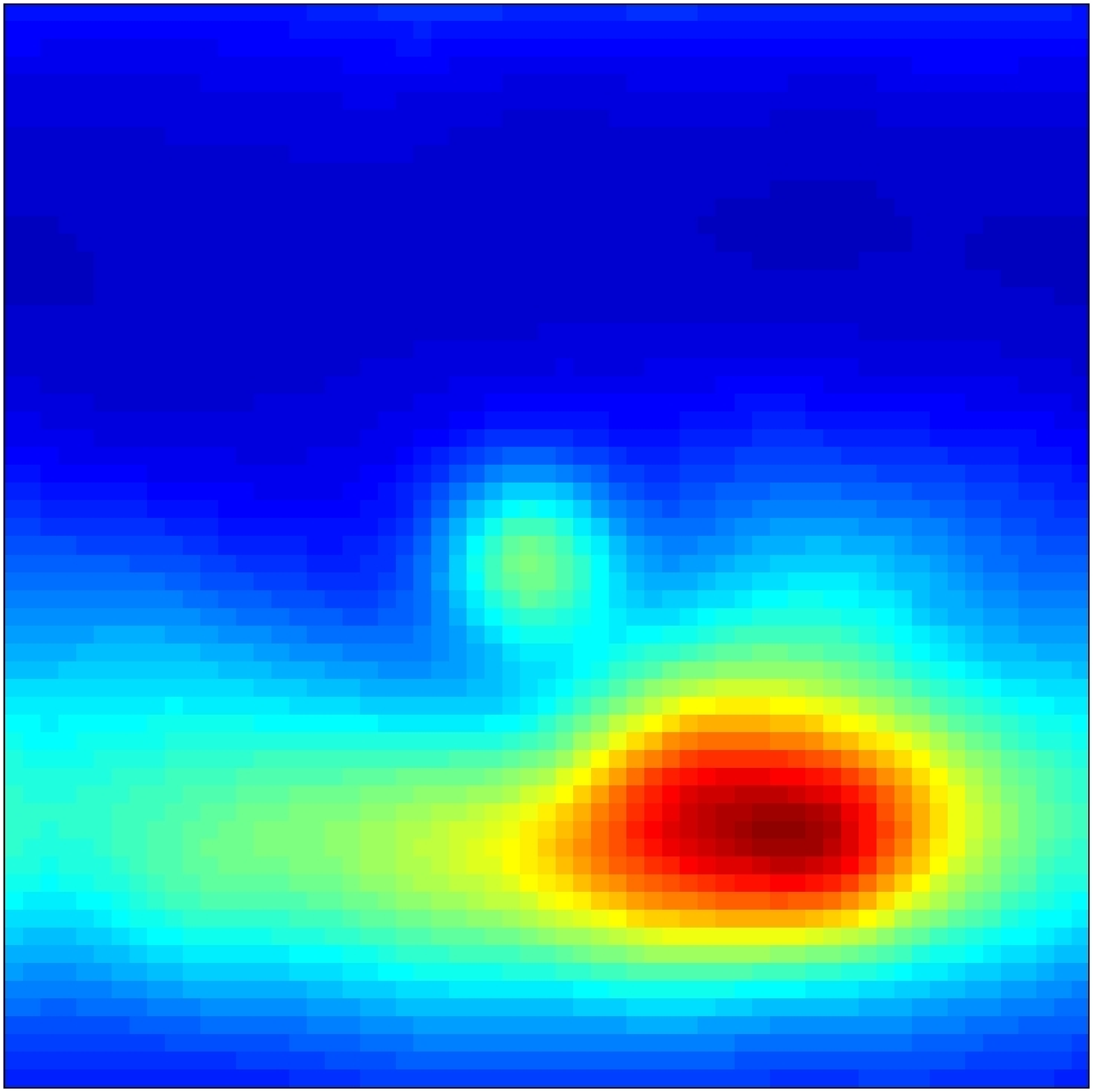}}
\quad
\subfigure[Thresholded SHW (constant threshold)]{\includegraphics[clip=,viewport=239 83 993 837,width=\recovplotwidth]{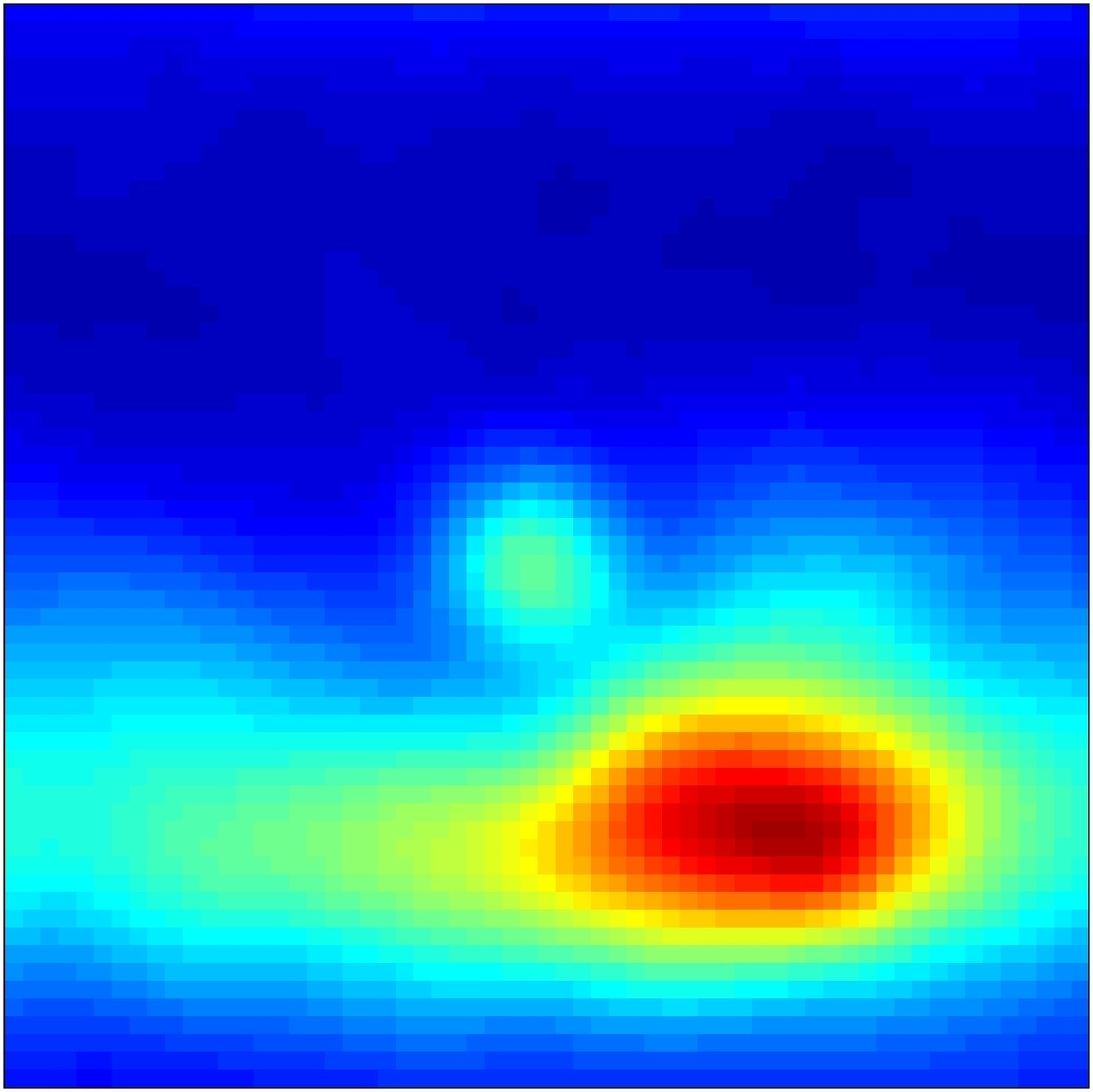}}
\quad
\subfigure[Thresholded SHW (annealing strategy)]{\includegraphics[clip=,viewport=239 83 993 837,width=\recovplotwidth]{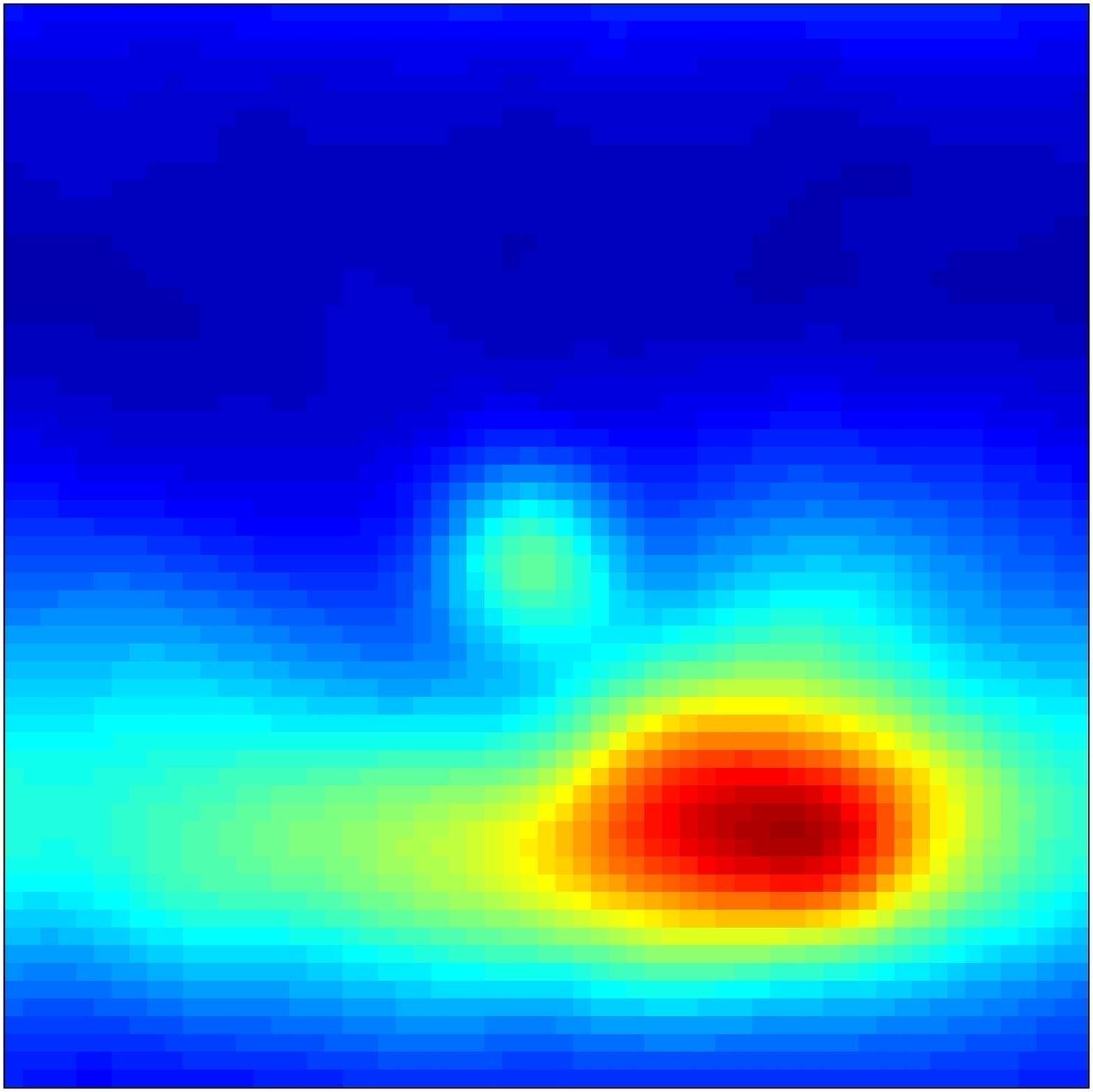}} 
}
\caption{Beam-modulated synchrotron intensity images for a
  $\sim\!\!20^\circ$ square patch.  The image shown in panel~(a) is
  constructed by projecting the full-sky beam-modulated intensity onto
  the tangent plane at the interferometer pointing direction defined by the
  coordinate system $\lxvect=(\lx,\mx)$ (as defined in
  \sectn{\ref{sec:fsi_coord}}).  The images shown in the remaining
  panels are constructed by simulating visibilities in the full-sky
  setting using all of the methods described in \sectn{\ref{sec:fsi}},
  followed by a standard inverse Fourier transform to recover the
  synthesised image.  The image in panel~(a) is expected to
  differ to the other images since full-sky contributions due to the
  large beam are incorporated when simulating visibilities, however a
  flat-patch approximation is assumed when synthesising images
  using the standard Fourier approach.  Nevertheless, all images are
  relatively similar, demonstrating and validating the use of the
  visibility formulations derived in \sectn{\ref{sec:fsi}} to simulate
  visibilities observed by an interferometer in the full-sky setting.
  The reconstructed images shown in panels (b), (c) and (d) are
  identical (to numerical precision) as expected, while the
  reconstructed images shown in panels (e) and (f) differ to these
  very slightly since some information is discarded when thresholding
  the wavelet coefficients in order to increase the speed of
  computations.}
\label{fig:recon_synchrotron}
\end{figure*}

In \tbl{\ref{tbl:comparison}} we compare the performance of the
methods used to compute visibilities in the simulations presented in
\fig{\ref{fig:recon_synchrotron}}.  The complexity of each method was
discussed in \sectn{\ref{sec:fsi}} and we collate these
complexities here and relate them to the maximum baseline of the
interferometer \blinemax.  For the exact full-sky interferometry
formalisms all coefficients are used, be they pixel values, spherical
harmonic coefficients or scaling and wavelet coefficients.  However,
the thresholded SHW strategies require only a small proportion of
wavelet coefficients due to the sparse representation of the
beam-modulated intensity function in the SHW basis.  Typically, less
than one percent of the wavelet coefficients can be retained while
introducing only minimal errors.  It is apparent that the annealed
thresholding strategy is slightly superior to the constant
thresholding approach, as one would expect since the annealing
approach is more likely to retain the detail coefficients of coarser
levels that inherently contain more information content.
%
The execution time tests presented in \tbl{\ref{tbl:comparison}} show
that the thresholded SHW methods are considerably faster than the other
methods at this resolution.  However, the true advantages of the
thresholded SHW methods will only become apparent at higher resolution
simulations.  At higher resolutions, the real, spherical harmonic and
naive SHW methods all scale like $\order(\blinemax{}^2)$.  The already
slow performance of these techniques and their poor scaling render
these methods computationally infeasible for high-resolution problems.
The thresholded SHW methods have much better scaling
properties (as discussed in \sectn{\ref{sec:fsi_shw_vis}}) and are
already considerably faster at this low-resolution, thus
rendering high-resolution simulations feasible.

\begin{table*}
\begin{minipage}{\textwidth}
  \caption{Comparison of the performance of the methods derived in
    \sectn{\ref{sec:fsi}} for simulating interferometric observations
    in the full-sky setting.  The results presented in this table
    correspond to the simulations illustrated in
    \fig{\ref{fig:recon_synchrotron}}. Execution time tests were
    performed on a laptop with a 2.2GHz Intel Core 2 Duo processor and
    2GB of memory.  }
\label{tbl:comparison}
\begin{center}
\begin{tabular}{lrrr} 
\toprule
\multicolumn{1}{l}{Method} &
\multicolumn{1}{r}{Complexity $\order(\blinemax{}^n)$} & 
\multicolumn{1}{r}{Coefficients retained\footnote{Note that the number
  of coefficients retained is only relevant for the thresholded SHW
methods; all other methods require all coefficients (be they pixel values, spherical
harmonic coefficients or SHW coefficients).}} & 
\multicolumn{1}{r}{Execution time} \\
\midrule
Direct quadrature                    & $n=2$        & 100.00\% & 207.6s\\
Spherical harmonic                   & $n=2$        & 100.00\% & 263.7s\\
Naive SHW                            & $n=2$        & 100.00\% & 238.9s\\
Thresholded SHW (constant threshold) & $n\lesssim1$ & 0.70\%   & 75.8s \\
Thresholded SHW (annealing strategy) & $n\lesssim1$ & 0.35\%   & 73.0s \\
\bottomrule
\end{tabular}
\end{center}
\end{minipage}
\end{table*}


\subsection{High-resolution simulated dust observations}
\label{sec:sim_dust}

In this section we simulate high-resolution observations of Galactic
dust emission made by an idealised interferometer.  Full-sky
visibilities are simulated using the annealing based thresholded SHW
method only since computational considerations preclude the
application of real and spherical harmonic representations, and the
annealing based SHW method was demonstrated in
\sectn{\ref{sec:sim_sync}} to be superior to the constant thresholding
method.  We subsequently refer to the annealing based SHW method for
simulating full-sky visibilities as the \emph{fast SHW method}.

\begin{figure}
\centering
\subfigure[Mollweide projection in global coordinates]{\includegraphics[width=70mm]{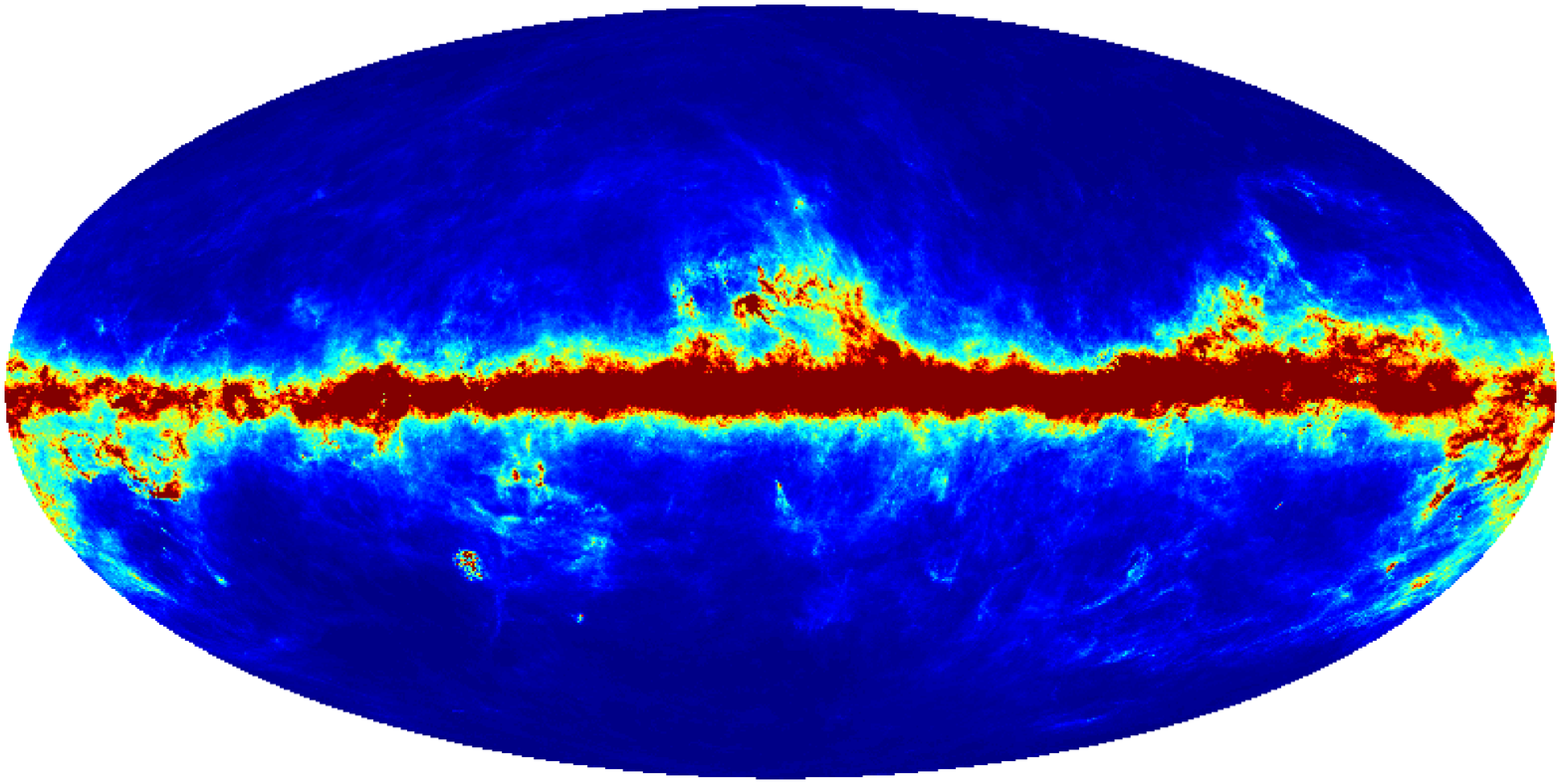}}
\subfigure[Globe in global coordinates]{\includegraphics[clip=,viewport=80 80 540 540,width=50mm]{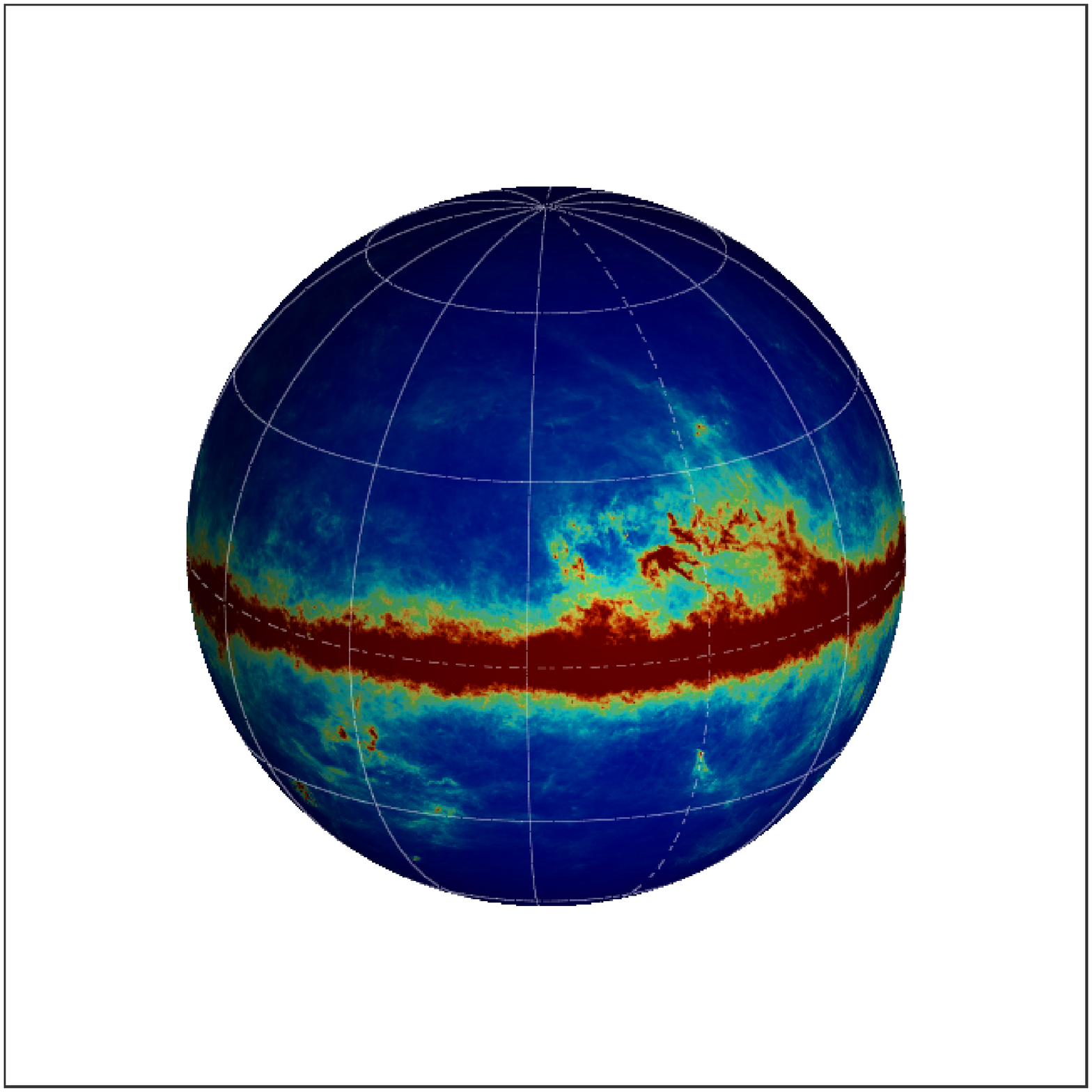}}
\subfigure[Globe in local coordinates]{\includegraphics[clip=,viewport=80 80 540 540,width=50mm]{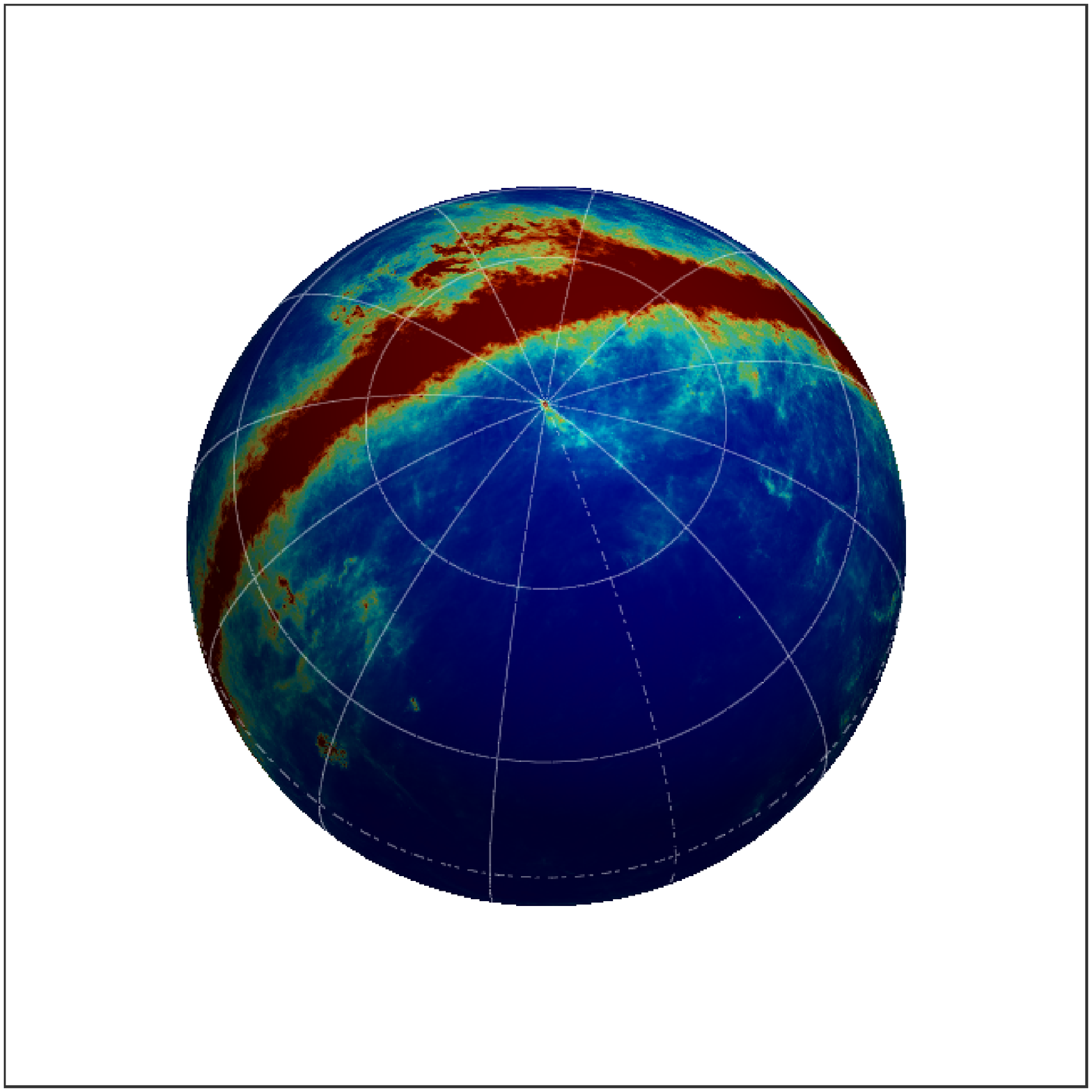}}
\subfigure[Zoomed globe in local coordinates]{\includegraphics[clip=,viewport=-10 -10 660 660,width=50mm]{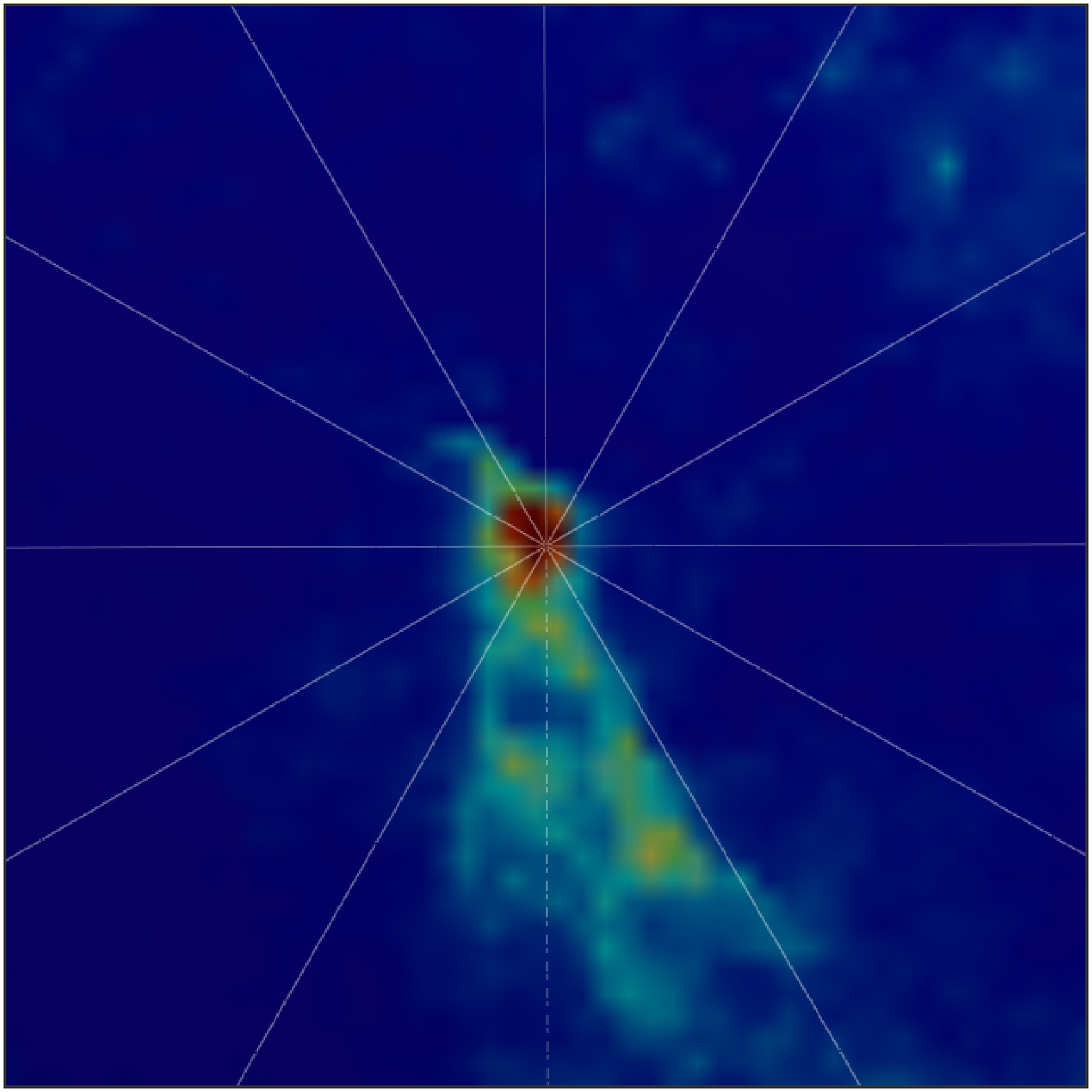}}

\caption{Full-sky 94GHz FDS map of predicted submillimeter and
  microwave emission of diffuse interstellar Galactic dust.  This map
  provides the full-sky background intensity map for our
  high-resolution simulated observations.  Panels (a) and (b) show the
  full-sky map in the global coordinate frame defined by Galactic
  coordinates.  We simulate observations of the source at
  $\saon=(\saa_0^\glob,\sab_0^\glob)=(108.0^\circ,0.0^\circ)$.  A
  rotation by $\roto^{-1}\equiv \rot(0,-\saa_0^\glob,-\sab_0^\glob)$
  is performed to convert the map to the local coordinate versions
  illustrated in panels (c) and (d), which are centred on the north
  pole.  }
\label{fig:lambda}
\end{figure}

The synchrotron map considered previously does not contain enough
high-frequency content for the high-resolution simulations performed
here.  For the background intensity map we therefore use the 94GHz FDS
map of predicted submillimeter and microwave emission of diffuse
interstellar Galactic dust \citep{finkbeiner:1999}.  This predicted
map is based on the merged Infrared Astronomy Satellite (IRAS) and
Cosmic Background Explorer Diffuse Infrared Background Experiment
(COBE-DIRBE) observations produced by \citet{schlegel:1998}.  An
undersampled version of the FDS map is available from \lambdaarch\ as a
\healpix\ sphere at resolution $\nside=512$.  The FDS map that we assume
as our full-sky background intensity map $\inteng(\sag)$, to be
observed by our idealised interferometer, is illustrated in
\fig{\ref{fig:lambda}}~(a) and (b) in global coordinates.  We simulate
observations of the extended source located at
$\saon=(\saa_0^\glob,\sab_0^\glob)=(108.0^\circ,0.0^\circ)$.  It is
necessary to convert the global intensity function to its local
coordinate version through a rotation by $\roto^{-1}$ in order to
compute visibilities.  The local coordinate intensity function
$\intenl(\sal)$ is illustrated in \fig{\ref{fig:lambda}~(c)} and (d).
The extended source to be observed is clearly visible at the north
pole of this map.  The interferometer beam used in these simulations is
given by a Gaussian with ${\rm FWHM_b}\simeq2.9^\circ$.

Given the mock data and configuration discussed above, we simulate
visibilities using the fast SHW method (\ie\ the annealing based
thresholded SHW approach).  We assume complete $uv$ coverage and
consider a baseline limit of $u_{\rm max} = 100$, and reconstruct an
image of size $\nimage = 20\times20$.
Following the approach performed in \sectn{\ref{sec:sim_sync}}, a
synthesised image is reconstructed from simulated visibilities simply
by taking the inverse Fourier transform.  The limitations of this
approach were discussed previously, however it is suitable for visual
verification of the simulations performed in this article.  Nyquist
sampling dictates that the synthesised image corresponds to a
$\sim5.7^\circ$ square patch.

\begin{figure}
\centering
\subfigure[Tangent plane image]{\includegraphics[width=50mm]{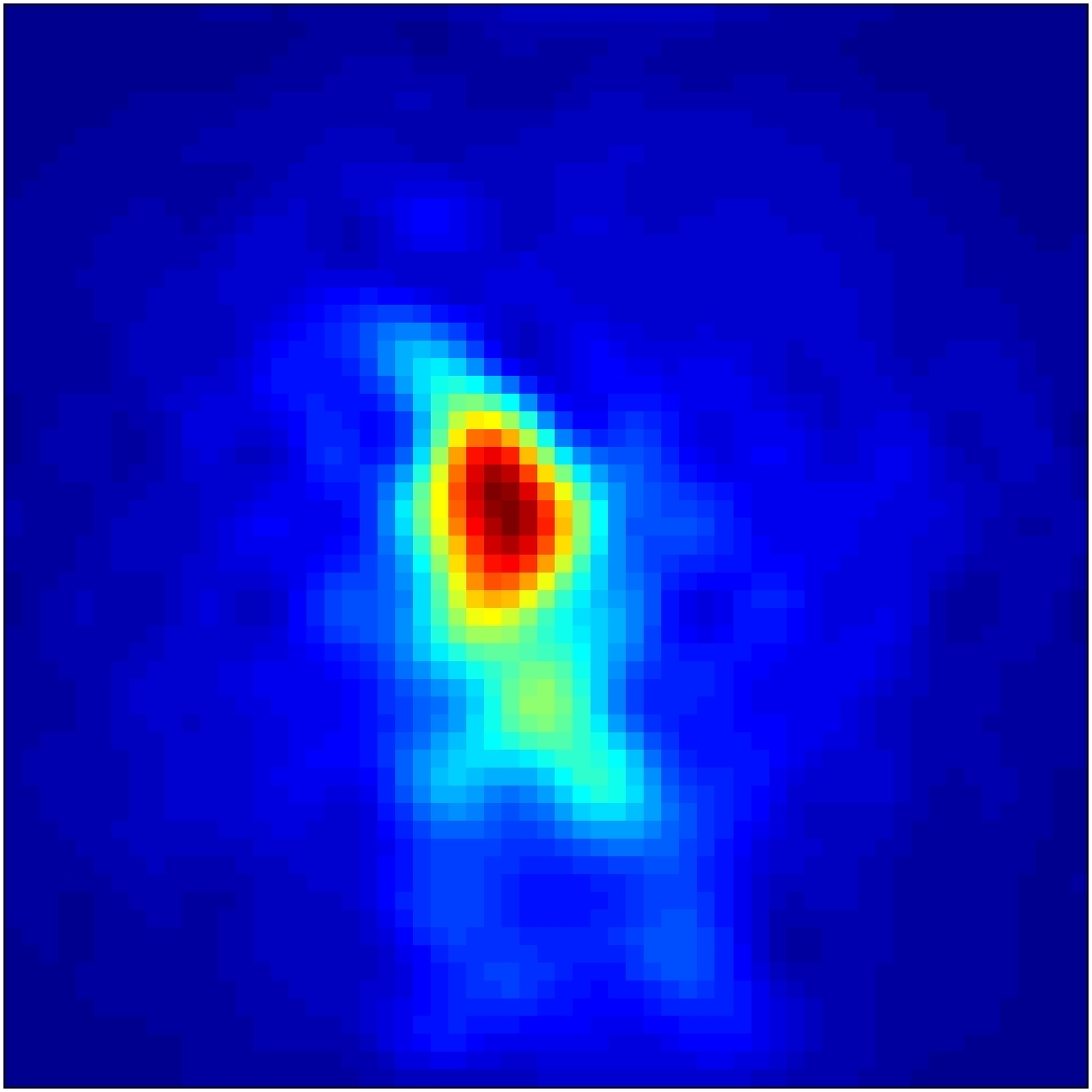}}
\subfigure[Simulated full-sky interferometric image]{\includegraphics[width=50mm]{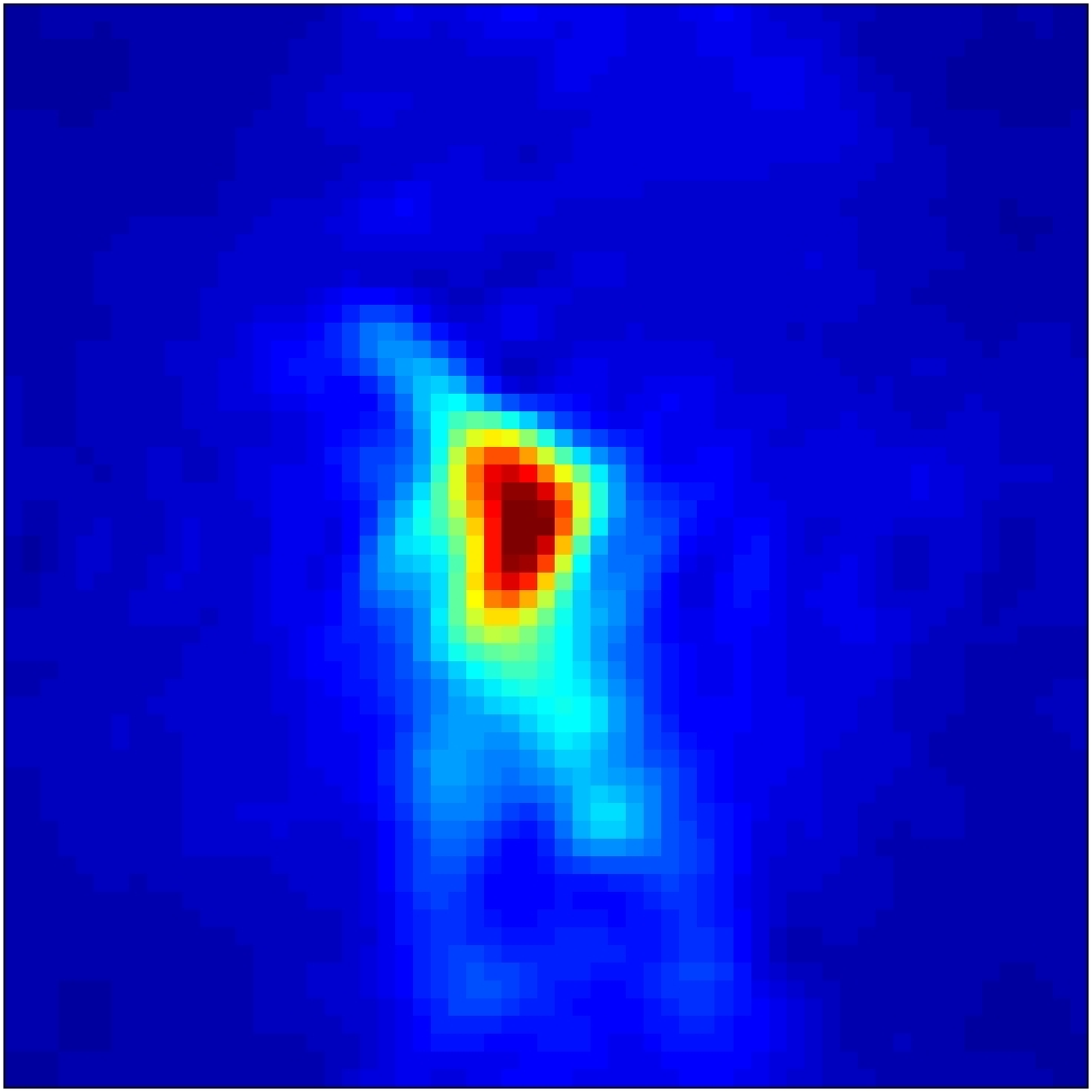}}
\caption{Beam-modulated Galactic dust intensity images for a
  $\sim\!\!5.7^\circ$ square patch.  The image shown in panel~(a) is
  constructed by projecting the full-sky beam-modulated intensity onto
  the tangent plane at the interferometer pointing direction defined
  by the coordinate system $\lxvect=(\lx,\mx)$ (as defined in
  \sectn{\ref{sec:fsi_coord}}).  The image shown in panel~(b) is
  constructed by simulating visibilities in the full-sky setting using
  the fast SHW method, followed by a standard inverse Fourier
  transform to recover the synthesised image.  The images in
  panels~(a) and (b) are expected to differ since full-sky
  contributions due to the large beam are incorporated when simulating
  visibilities, however a flat-patch approximation is assumed when
  synthesising the image using the standard Fourier approach.
  Furthermore, the fast SHW method itself introduces some small error
  when discarding those wavelet coefficients with minimal information
  content.  Nevertheless, the two images are relatively similar,
  demonstrating and validating the use of the fast SHW method to
  simulate full-sky visibilities in a high-resolution setting.}
\label{fig:recon_lambda}
\end{figure}

The direct projection of the full-sky beam-modulated intensity
function onto the tangent plane at the interferometer pointing
direction is illustrated in \fig{\ref{fig:recon_lambda}~(a)}.  The
beam-modulated intensity image reconstructed from the full-sky
simulated visibilities is illustrated in
\fig{\ref{fig:recon_lambda}~(b)} (upsampled to a 60$\times$60 pixel
image for visualisation).
In this simulation visibilities are computed using 0.023 percent of
the wavelet coefficients of the beam-modulated intensity function only.
Again, one expects the tangent plane and reconstructed images to
differ slightly since full-sky contributions are incorporated when
simulating visibilities, however a flat-patch approximation is assumed
when synthesising the image using the standard Fourier approach.
Furthermore, the fast SHW method itself introduces some small error
when discarding those wavelet coefficients with minimal information
content and the interferometer baseline limit also restricts the
high-frequency content of the reconstructed image.  Nevertheless, the
tangent plane and reconstructed image are in close agreement,
demonstrating and validating the application of the fast SHW approach
to simulate full-sky visibilities observed by an interferometer in a
high-resolution setting.

The high-resolution simulations performed here using the fast SHW
method required an execution time of 289.6s.  Based on the scaling
relationships of the real and spherical harmonic space methods, and
their execution times on the low-resolution simulations performed in
\sectn{\ref{sec:sim_sync}}, we estimate the execution time of these
methods when applied to these high-resolution simulations to be
$\sim3000$s.  As the resolution of the problem (\ie\ the baseline
limit $u_{\rm max}$) and the number of visibilities to be computed
(\ie\ $\nimage$ in this notation) increase, this difference will
become even more marked.  The fast SHW method is therefore essential
to compute realistic simulations of visibilities when incorporating
full-sky contributions.  In future, we intend to parallelise our
implementation of the fast SHW method and apply it to produce
realistic simulations of full-sky interferometer observations,
including incomplete $uv$ coverage and other more realistic
assumptions.

\section{Conclusions}
\label{sec:conclusions}

Next generation interferometers will have very large fields of view,
which poses a number of imaging challenges.  The usual interferometric
approach to simulate visibilities or reconstruct images is based on
standard Fourier imaging, which relies on a tangent plane
approximation that is valid only for a small field of view.  This
approach is inappropriate for the wide fields of view proposed for the
next generation of interferometers.  In this article we have
formulated interferometry in the full-sky setting, incorporating
contributions over the entire sky to ensure that contamination due to
wide sidelobes of primary beams is not neglected.
Full-sky interferometry formulations have been developed in real,
spherical harmonic and SHW spaces.  The SHW formulation proves the
most advantageous approach due to the efficient representation of the
spatially localised high-frequency content typical of primary beams and sky
intensity functions in the wavelet basis.  A corresponding fast SHW
method was developed to simulate full-sky visibilities.  The fast SHW
method exploits the sparsity of the wavelet representation by
discarding all wavelet coefficients that contain minimal information
content.  A quadratic annealing based thresholding strategy was
developed to determine those wavelet coefficients that may be safely
discarded.  The resulting fast SHW method has superior computational
scaling properties and may be performed at a substantially lower
computational cost than the real and spherical harmonic space methods
for simulating visibilities when including full-sky contributions.

The primary focus of this article is the use of these full-sky
interferometry representations to simulate visibilities, however we
also briefly discussed implications for the reconstruction of images
on wide fields of view.  Although it is possible in theory to
reconstruct full-sky images in the spherical harmonic representation,
this requires full sampling of the visibility function in $\reals^3$
and hence is not likely to be well posed in practice.  However, image
reconstruction on wide fields of view is likely to be well posed in
SHW space and we outlined a preliminary method to perform wide field
image reconstruction.  The use of SHWs for image reconstruction is an
interesting application in its own right and we intend to develop
these ideas further and to evaluate their effectiveness in a future
work. 

To demonstrate the application of our full-sky interferometry
representations for simulating visibilities, and to compare the
various methods, we simulated low-resolution full-sky interferometric
observations of synchrotron emission using all of the methods.  The
real, spherical harmonic and naive SHW methods recovered identical
images of the extended source observed by our idealised
interferometer, where the naive SHW method does not exploit sparse
representations in the wavelet basis.  The fast SHW method does
exploit sparse representations, introducing a small error when
discarding those wavelet coefficients that contain minimal information
content, but consequently the speed of computations in increased
substantially.  Typically less than one percent of wavelet
coefficients are retained in these low-resolution simulations,
reducing the execution time of simulations by a factor of
approximately three compared to the other methods (while introducing
only minimal errors in reconstructed images).  However, the true
advantages of the fast SHW method only become apparent on
high-resolution simulations due to its superior scaling properties.

The already slow performance of the real and spherical harmonic space
approaches to simulating full-sky visibilities, and their poor scaling
properties, render these methods computationally infeasible on higher
resolution problems.  Computing full-sky visibilities using the fast
SHW method adapts to the extremely sparse representation of the beam
and sky intensity function in the wavelet basis, thus easing the
computational burden of simulations to the extent that they are
rendered computationally feasible at high-resolutions.
Using this method high-resolution interferometric observations of
diffuse interstellar Galactic dust were simulated, demonstrating and
validating the use of the fast SHW method for simulating visibilities
in a high-resolution setting.  In this example only 0.023 percent of
wavelet coefficients were retained and the execution time of
simulations was estimated to be greater than an order of magnitude
faster than simulations based on the real or spherical harmonic space
full-sky interferometry formulations.

Now that it is possible to simulate the visibilities observed by an
interferometer when including contributions over the entire sky, a
number of related studies may be performed.  Firstly, we intend to
parallelise our implementations and develop more realistic simulations
of full-sky interferometer observations, including incomplete $uv$
coverage and other more realistic assumptions.  Using these
simulations, we then intend to study the effect of contamination in
realistic interferometric observations due to wide sidelobes of
primary beams.  Secondly, our full-sky interferometry formulation
allows one to simulate visibilities for all values of
$\blinel=(u,v,w)$, and not only values of $\blinel$ where $w=0$, as is
the case with the standard Fourier transform approach.  Such
simulations allow one to evaluate the performance of the faceting
\citep{cornwell:1992,greisen:2002} and $w$-projection
\citep{cornwell:2005} approaches to wide field image reconstruction on
simulations where the ground truth is known.  Thirdly, we have
highlighted an interesting alternative approach to wide field image
reconstruction based on the SHW representation that warrants further
study.  We intend to pursue all three of these applications in future
work.  The ability to perform realistic high-resolution simulations of
interferometric observations that include full-sky contributions,
afforded by our fast SHW formulation, allows many new studies to be
performed that will be important to the design of next generation
interferometers and to the development of algorithms to analyse the
data generated by these instruments.

\section*{Acknowledgements}

We thank Paul Alexander and Mike Hobson for useful comments.  Some of
the results in this paper have been derived using the \healpix\
package \citep{gorski:2005}.  We acknowledge the use of the
\lambdaarchtext\ (\lambdaarch).  Support for \lambdaarch\ is provided
by the NASA Office of Space Science.

\bibliographystyle{mymnras_eprint}
\bibliography{bib}

\label{lastpage}
\end{document}